%% file: main.tex
\documentclass[review]{elsarticle}

\usepackage{xspace}
\usepackage{makecell}
\usepackage{caption}
\usepackage{subcaption}
\usepackage{url}
\usepackage{mdframed}
\usepackage{totcount}
\usepackage{multirow}
\usepackage{amssymb}
\usepackage{booktabs}
\newtotcounter{findings}
\newcommand\findings{\stepcounter{findings}\arabic{findings}}

\AtBeginDocument{%
  \providecommand\BibTeX{{%
    \normalfont B\kern-0.5em{\scshape i\kern-0.25em b}\kern-0.8em\TeX}}}

\begin{document}
\title{A Qualitative and Quantitative Analysis \\ of Container Engines}

\author{Luciano Baresi, Giovanni Quattrocchi, Nicholas Rasi}
\address{Dipartimento di Elettronica, Informazione e Bioingegneria\\Politecnico di MIlano, Milan, Italy\\\url{{name.surname}@polimi.it}}









\def\CLEAN{1}
\if\CLEAN0
\newcommand{\nb}[3]{
	\fcolorbox{black}{#3}{\bfseries\sffamily\scriptsize#1}
	{\sf\small \textit{#2}}
}
\newcommand\red[1]{\textcolor{red}{#1}}
\newcommand\blue[1]{\textcolor{blue}{#1}}
\else 
\newcommand{\nb}[3]{}
\newcommand\red[1]{\textcolor{black}{#1}}
\newcommand\blue[1]{\textcolor{black}{#1}}
\fi

\newcommand\luc[1]{\nb{LB}{#1}{red}}
\newcommand\gio[1]{\nb{GQ}{#1}{orange}}
\newcommand\nic[1]{\nb{NR}{#1}{green}}

\begin{abstract}
    Containerization is a virtualization technique that allows one to create and run executables consistently on any infrastructure. Compared to virtual machines, containers are lighter since they do not bundle a (guest) operating system but they share its kernel, and they only include the files, libraries, and dependencies that are required to properly execute a process. In the past few years, multiple container engines (i.e., tools for configuring, executing, and managing containers) have been developed ranging from some that are ``general purpose'', and mostly employed for Cloud executions, to others that are built for specific contexts, namely Internet of Things and High-Performance Computing.
    Given the importance of this technology for many practitioners and researchers, this paper analyses six state-of-the-art container engines and compares them through a comprehensive study of their characteristics and performance. The results are organized around \total{findings} findings that aim to help the readers understand the differences among the technologies and help them choose the best approach for their needs.
\end{abstract}




\maketitle

\section{Introduction}
\label{sec-introduction}
\input{1_intro}

\section{Identified Engines}
\label{sec-framework}
\input{2_requirements}

\section{Qualitative analysis} 
\label{sec-features}
\input{4_features}

\section{Quantitative Evaluation}
\label{sec-evaluation}
\input{5_evaluation}

\subsection{HPC Evaluation}
\label{sec-hpc-results}
\input{6_hpc}

\section{Discussion}
\label{sec-discussion}
\input{6-5._discussion}

\vspace{0.2cm}

\section{Related Work}
\label{sec-related}
\input{7_related}

\section{Conclusions and Future Work}
\label{sec-conclusion}
\input{8_conclusion}

\bibliographystyle{elsarticle-harv}
\bibliography{bib}



\end{document}

%% file: 1_intro.tex
Containerization~\cite{containers1} is a widely adopted virtualization technology that allows one to create lightweight executables that can run consistently on any infrastructure.  They exploit a shared operating system (OS) and package executables along with their required dependencies (e.g., code, configuration files, and libraries) in a standard format. They are lighter, faster to boot, and scale better than virtual machines (VM) because they share a host OS kernel and only virtualize userspaces~\cite{containers2,baresi2020cocos}. Originally conceived for cloud-native applications~\cite{pahl2015containerization}, they are now the de-facto compute unit in many contexts ---for example, Internet of Things (IoT)~\cite{celesti2016exploring} and High-Performance Computing (HPC)~\cite{higgins2015orchestrating} solutions. 
Containers mainly exploit the features of the Linux kernel, but there are also a few implementations for other operating systems~\cite{windows_container}.  

A \textit{container engine} supplies a set of tools to create and manage containers. It provides means to create \textit{container images}, that is, container blueprints that embed all the necessary files needed at runtime: a \textit{container} (or \textit{container instance}) is then a standard process that instantiates a container image. Multiple containers can be created from a single image.  Images are usually defined through dedicated languages (e.g., Dockerfile~\cite{Dockerfile}) and built using \textit{image builders}. The \textit{container runtime} allows one to start and manage containers from images, configure the kernel, and initialize the process that the container wraps. Container engines are usually connected to a public (or private) \textit{registry}, where users can upload or download container images, and can also provide \textit{orchestration tools} to schedule, connect, and oversee the execution of multiple containers running on a cluster of physical or virtual machines.
In the last few years, numerous container engines have been presented and adopted by practitioners. Each approach is dedicated to an execution environment and provides different features, security guarantees, and performance. 

This paper aims to shed some light on these technologies and help the readers understand the technical differences among the approaches and choose the best technology for their needs. Given the diverse engines that exist today, we selected three reference domains: cloud computing, IoT (Internet of Things) solutions, and HPC (High Performance Computing) systems, to categorize the different engines, understand the particular requirements, and emphasize the specificity of studied engines. For each domain, we retrieved the two most famous container engines, according to GitHub. We then analyzed Docker~\cite{docker_website}, which played a key role to made containers become popular, and Podman~\cite{podman_website} as cloud-native solutions, Charliecloud~\cite{charliecloud_website} and Singularity~\cite{singularity_website} as HPC-specific engines, and we only found balenaEngine~\cite{balena_website} as container engine for IoT systems. We decided to also add Sarus~\cite{sarus_website}, as third HPC-specific engine since it is a promising  solution and it is the only one that is fully developed in academia.

The proposed analysis is both qualitative, since we confront offered features, and quantitative, since we compared their performance on a comprehensive set of benchmarks. Obtained results are summarized in \total{findings} findings that highlight the characteristics, similarities, and differences of considered engines. 

The rest of this paper is organized as follows. Section~\ref{sec-framework} describes the selection process and the container engines we selected. Section~\ref{sec-features} presents the feature-wise comparison of the technologies. Section~\ref{sec-evaluation} shows our empirical evalation
\blue{, while Section~\ref{sec-discussion} discusses the results}. Section~\ref{sec-related} illustrates some related works and Section~\ref{sec-conclusion} concludes the manuscript.

%% file: 2_requirements.tex
Containers are not only used in the cloud anymore; other computing contexts exploit them and appreciate their characteristics.  Docker was the first engine that gained traction among practitioners, and other solutions  have been developed over the last years to address particular requirements and provide more appropriate engines. The characteristics of cloud-based engines, which are often common to all solutions, along with the specific needs that come from HPC and IoT applications, explain why we decided to identify three groups: cloud-specific, or general-purpose solutions, and HPC- and IoT-centric engines. We then explain how we selected the six engines by considering their popularity on GitHub, and we sketch their main characteristics.

\subsection{Contexts }
\label{subsec-features_reqs}

Containers are used to wrap applications in lightweight, isolated, and self-contained executables that are easily portable among different execution environments. They allow developers to create software in a local machine without worrying about compatibility issues that may arise when migrating to production environments. 

Containers have gained a lot of traction in \textbf{cloud} environments because they increase the speed of release cycles (i.e., more agility) and reduce the number of operations required (i.e., lower costs)~\cite{10.1145/3106237.3106270}. 
Moreover, oftentimes modern cloud-based applications are developed by following the microservices architecture~\cite{dragoni2017microservices} where, the application is structured in a set of loosely-coupled, ``small'' services that can be developed, tested, and deployed independently. In such a setup, each service can be packaged as a lightweight container to isolate their execution and management without adding any significant overhead.  Containers can also be used to package off-the-shelf monolithic applications to add portability and isolation.

Novel use cases for containers have emerged recently. Bentaleb et al.~\cite{bentaleb2022containerization} have noticed that in addition to cloud computing, HPC (High-Performance Computing) and IoT (and, similarly, Edge and Fog infrastructures) could benefit from the adoption of containers, but they pose some additional requirements. 

\blue{
\textbf{HPC} systems consist of nodes that are equipped with specialized hardware, networking, and storage solutions designed for highly parallel computations. These systems often have a large number of CPUs and GPUs, and the computation on these devices is typically managed by a workload manager. In this context, managing configurations can be difficult due to the complexity of the platform, the need to integrate with the workload manager, the requirement for specialized libraries (e.g., MPI\footnote{https://www.mpi-forum.org}), and the differences between development environments (i.e., standard computers and the HPC center that hosts the execution).}

\blue{
Containers can help simplify the management of dependencies and libraries, and increase the portability of applications by abstracting away differences between execution environments. This enables the integration of HPC applications into continuous integration pipelines~\cite{10.1145/3219104.3219147}, which have been shown to improve software quality and speed up application evolution~\cite{9374092}.}

\blue{
Furthermore, containers can help isolate resources among different processes without introducing a significant overhead. This is particularly relevant in highly distributed and concurrent setups like HPC infrastructures, where resource contention can significantly reduce overall system performance~\cite{8466019}. With containers, each process is isolated and can be configured to use a set of resources exclusively (which will prevent resource contention by design) or to share them with other containers by defining the priority of each container in the event of resource saturation. The combination of reliable dependency management and deterministic resource allocation mechanisms provided by containers enables more deterministic and reproducible executions, which can help both practitioners and researchers design better HPC-based services and understand their performance.}

\blue{
In addition to the challenges posed by the complexity of HPC systems and the need to manage dependencies and configurations, HPC centers are also multi-tenant and multi-user systems, which makes security a critical concern. In order to protect against security threats such as malware and unauthorized access, it is important for processes to be executed without privileged permission (rootless executions). This helps to prevent processes from gaining access to sensitive resources or from making unauthorized changes to the system. }

\blue{\textbf{IoT} nodes are made up of a wide variety of different hardware architectures, configurations, and operating systems, which can make it challenging to deploy and manage applications across this heterogeneous environment. In addition, IoT devices often have limited resources such as memory, storage, and processing power, which can make it necessary to use lightweight and highly optimized solutions.}

\blue{Containers can help address these challenges by increasing the portability of applications across devices with different hardware architectures (e.g., ARM or x86), and by enabling the optimization of resource usage. Containers can also help isolate concurrent processes running on the same device and manage available resources, which is important to prevent resource saturation and to carefully allocate resources to each process. This is particularly important in IoT environments where the number of devices may be large and dynamic, and where the demand for resources may vary over time. By using containers, it is possible to easily increase or decrease the number of instances of an application running on IoT devices, providing a flexible and scalable solution.}

\blue{Finally, the exchange of data between IoT nodes can often involve sensitive or privacy-critical information, and it is important for container engines to prioritize security in order to protect this data and prevent the compromise of devices. To prevent data from being intercepted and accessed by unauthorized parties, it is important to encrypt data as it is transmitted between IoT devices. In addition, container engines should implement measures to ensure that data is not tampered with or modified by unauthorized parties.}

As described herein, cloud computing, HPC, and IoT systems pose different challenges and there is no one-size-fits-all solution that can easily accommodate all of them. As we will discussed in the rest of the paper, the state-of-the-art offers container engines that are ``general-purpose'' and mainly used in cloud setups and others that are specialized to a single context to better address its main issues.

\subsection{Selection process}

To select the most important container engines in each context (Cloud, IoT, and HPC), we used a semi-automated pipeline\footnote{The results of each step can be found at: \url{https://github.com/deib-polimi/container-engines-selection}}: i) GitHub queries, ii) automated filtering, and iii) manual labelling. 

We started retrieving a set of GitHub repositories that may provide a container engine dedicated to a given context by issuing three dedicated queries (\textit{container + $<$context$>$}) to GitHub, where \textit{$<$context$>$} was \textit{``''}, to mean ``standard'' cloud solutions, and \textit{hpc} and \textit{iot} for the other two cases.
The search was restricted to either README files or repository descriptions, and we set the number of GitHub stars of an acceptable repository to be equal to or greater than $100$. We sorted the results by stars and we analyzed the first $1000$ results for each query. At the end of the first step, we obtained $1000$, $90$, and $385$ repositories for cloud-, HPC-, and IoT-focused repositories, respectively. 

As additional refinement, we automatically filtered out some of the results not related to container engines. In particular, we removed all the repositories whose descriptions (lower-cased and without punctuation and symbols) did not contain any of the following keywords: \textit{cloud}, \textit{container}, \textit{iot}, \textit{internet of things}, \textit{hpc}, and \textit{high performance computing}. After this step, we remained with $104$, $21$, and $49$ repositories, respectively.

As final step, we manually scanned the results to properly identify container engines. We used four labels to characterize  each repository: \texttt{ENGINE} (a container engine), \texttt{RUNTIME} (a low-level container engine component) , \texttt{TOOL} (a tool related to containers but not a container engine or runtime), and \texttt{N/A} (a non-relevant element).
 
Among the $104$ repositories retrieved for cloud container engines, we tagged $4$ with label \texttt{ENGINE}, $3$ with label \texttt{RUNTIME}, and, $46$ and $51$ with labels \texttt{TOOL} and \texttt{N/A}, respectively. The four identified container engines are \textit{Docker}\footnote{Each engine has its own repository. The core functionality of Docker is now included in the Moby project: \url{https://github.com/moby/moby}. Podman is available from: \url{https://github.com/containers/podman}. RKT is available from: \url{https://github.com/rkt/rkt}, but it was discontinued in 2020 as reported at \url{https://github.com/rkt/rkt/issues/4024}. Pouch is available from: \url{https://github.com/alibaba/pouch}, containerd from: \url{https://github.com/containerd/containerd}, run-c from: \url{https://github.com/opencontainers/runc}, cri-o from: \url{https://github.com/cri-o/cri-o}, Charliecloud from: {\url{https://github.com/hpc/charliecloud}, Singularity from: \url{https://github.com/sylabs/singularity}, Apptainer from: \url{https://github.com/apptainer/apptainer}, balenaEngine from: \url{https://github.com/balena-os/balena-engine}, and Sarus from: \url{https://github.com/eth-cscs/sarus}.}

} (62.7k stars), \textit{Podman} (13.2k stars), \textit{RKT} (8.8k stars) and \textit{Pouch} by Alibaba (4.4k stars); while the three discovered container runtimes are \textit{containerd} (10.6k discovered), \textit{run-c}  (9.0k stars), and \textit{cri-o} (3.9k stars).
In the context of HPC, we obtained $3$ engines, no runtimes, $6$ tools, and $12$ repositories marked as  \texttt{N/A}. The three identified container engines are \textit{Charliecloud} (238 stars), \textit{Singularity} by Sylabs (214 stars), and \textit{Apptainer} (196 stars). The last two engines share the same core functionality being originated from the same project (i.e., Singularity has been recently renamed Apptainer).
Finally, for IoT we discovered one single engine,  \textit{balenaEngine} (586 stars), no runtime, $6$ tools, and $42$ repositories marked as \texttt{N/A}.

\begin{table}[t]
    \centering
	\begin{tabular}{ccccc}
		\toprule
		  \textbf{Name} & \textbf{FRY} & \textbf{\#R} & \textbf{LOC} & $\star$ \\
		\midrule
		Docker        & 2013 & 123 & 2.331.244 & 62.7k \\
		Podman        & 2017 & 96 & 971.647 & 13.2k \\
		Charliecloud  & 2015 & 25 & 8.071 & 238 \\
		Singularity   & 2015 & 78 & 62.819 & 214 \\
		Sarus         & 2018 & 10 & 16.890 & 78 \\
		balenaEngine  & 2015 & 127 & 1.051.024 & 586 \\
		\bottomrule
	\end{tabular}
	\caption{Container engines.}
	\label{tab:impl}
\end{table}

We selected the two container engines (if available) with the highest amount of stars from each group, and selected five engines: Docker, Podman, Charliecloud, Singularity, and balenaEngine. 
Because of its characteristics, we also added Sarus\footnote{\url{https://github.com/eth-cscs/sarus}} (78 stars), an emerging solution completely developed in academia, and optimized for HPC.

For each selected engine, Table~\ref{tab:impl} reports the year of the first release ($FRY$), the number of releases ($\#R$) until now, the size of the project in lines of code ($LOC$), and the amount of GitHub stars retrieved by our queries ($\star$), to estimate the adoption rate and its reputation.

%% file: 4_features.tex

\begin{table}[t]
    \small
    \centering
    \setlength{\tabcolsep}{2.3pt}

	\begin{tabular}{lcccccc}
		\toprule
		  \textbf{Feature} & \textbf{Doc}  & \textbf{Pod} & \textbf{Cha} & \textbf{Sin} & \textbf{Sar} & \textbf{bal} \\
		\midrule
		\multicolumn{7}{c}{\textit{\textbf{Compatibility}}}  \\
		\textit{OCI compliance} & \checkmark & \checkmark
& $\sim$  & \checkmark & \checkmark & \checkmark \\ 
\textit{Support for multiple host operating systems} & \checkmark & \checkmark  & \checkmark & \checkmark & \checkmark & $\times$ \\
 \midrule
 \multicolumn{7}{c}{\textit{\textbf{Image management}}}  \\
 \textit{Built-in image builder} & \checkmark & \checkmark
& \checkmark & \checkmark & $\times$ & \checkmark \\
    \textit{Images from Dockerfiles/Containerfiles} & \checkmark & \checkmark
& \checkmark & \checkmark & $\times$ & \checkmark \\  
    \textit{Images from SDFs} & $\times$ \ & $\times$
& $\times$ & \checkmark & $\times$ & $\times$ \\ 
        \textit{Images from git repositories} & \checkmark & \checkmark
& $\times$ & $\times$ & $\times$ & \checkmark \\  
    \textit{Images from archives} & \checkmark & \checkmark
& \checkmark & $\times$ & $\times$ & \checkmark \\  
        \textit{Public/private registries} & \checkmark & \checkmark
& \checkmark & \checkmark & \checkmark & \checkmark \\
        \textit{Support for Docker/OCI images} & \checkmark & \checkmark
& $\sim$ & \checkmark & \checkmark & \checkmark \\
        \textit{Support for SIF} & $\times$ & $\times$
& $\times$ & \checkmark & $\times$ & $\times$ \\
\midrule
 \multicolumn{7}{c}{\textit{\textbf{Optimizations}}}  \\
 \textit{Layered images} & \checkmark & \checkmark
& \checkmark & $\times$ & \checkmark & \checkmark \\
\textit{Differential image updates} & $\times$ & $\times$
& $\times$ & $\times$ & $\times$ & \checkmark \\
\textit{On the fly image extraction} & $\times$ & $\times$
& $\times$ & $\times$ & $\times$ & \checkmark \\
\textit{Image compression} & $\times$ & $\times$
& $\times$ & \checkmark & \checkmark & $\times$ \\ 
\midrule
 \multicolumn{7}{c}{\textit{\textbf{Containers executions}}}  \\
 \textit{Detached executions} &  \checkmark  &  \checkmark 
& $\times$ & \checkmark & $\times$ & \checkmark  \\
\textit{Dynamic Resource Management} & \checkmark & $\times$
& $\times$ & \checkmark & $\times$ & $\times$ \\
\textit{Built-in orchestration capabilities} & \checkmark & $\times$
& $\times$ & $\times$ & $\times$ & \checkmark \\
\textit{Support for workload managers} &  $\times$  &  $\times$ 
 & \checkmark & \checkmark & \checkmark & $\times$  \\
\textit{Support for GPUs} & \checkmark & $\times$
& $\sim$ & \checkmark & \checkmark & \checkmark \\
\textit{Explicit support for MPI libraries} & $\times$ & $\times$
& $\sim$ & \checkmark & \checkmark & $\times$ 
 \\\midrule 
  \multicolumn{7}{c}{\textit{\textbf{Security}}}  \\ 
\textit{Signed container images} & \checkmark & \checkmark
& $\times$ & \checkmark & $\times$  & \checkmark \\
\textit{Rootless executions} & $\sim$ & \checkmark
& \checkmark & \checkmark & $\times$ & $\times$ \\
\textit{Daemonless execution model} & $\times$ & \checkmark
& \checkmark & \checkmark & \checkmark & $\times$ \\
\textit{VPN support} & $\times$ & $\times$
& $\times$ & $\times$ & $\times$ & \checkmark \\
		\bottomrule
	\end{tabular}
	 \caption{Qualitative comparison: \checkmark available, $\sim$ partially available, and $\times$ not available.}
	\label{tab:features}
\end{table}

All six container engines are open source and their development is active. Most of the implementations are written in Go, but Charliecloud and Sarus: the former is written using a mix of shell scripts and C code, the latter is written in C++. Even if we have identified three groups, each solution, with the exception of balenaEngine that requires a dedicated OS, can be used in any context due to the natural portability of containers (at least on Linux-based systems). 

Docker~\cite{docker_website} was the first popular engine, and it is still probably the most famous one. It provides a complete production platform for developing, distributing, securing, and orchestrating container-based solutions. It is easy to use, well established among developers, and its general-purpose nature fits almost any software project.  Docker exploits \textit{containerd}~\cite{containerd_website} for managing and running containers, which in turn uses \textit{run-c}~\cite{runc_website} as low-level container runtime for container creation.
Podman~\cite{gantikow2020rootless} is younger compared to Docker. It is based on library \textit{libpod}, from the same developers, that is used for managing the entire container lifecycle. It uses, like Docker, the \textit{run-c} container runtime for instantiating containers.

Charliecloud~\cite{charliecloud_paper1} focuses on the management of containers on HPC frameworks by simplifying the packaging and transmission of HPC applications, and is the lightest solution among the analyzed ones (i.e., only a few thousands of LOC).
Singularity~\cite{singularity_paper1} is a container engine optimized for the computation of HPC workloads. It can also be used in cloud deployments and supports Docker images and registries. From version 3.1.0, Singularity is fully OCI compliant, and supports both OCI specifications: image-spec and runtime-spec.
Sarus~\cite{sarus_paper} is the most recent project, with the lowest number of releases, and has been designed to run containers in the HPC context. It leverages \textit{run-c} as container runtime, and provides an extensible runtime to support current and future custom hardware while achieving native performance. 

balenaEngine~\cite{balena_website} is based on the Moby Project~\cite{moby_website}, a framework provided by Docker to assemble specialized container systems, and offers a complete platform for deploying IoT applications using containers. 

Table~\ref{tab:features} summarises our analysis and organizes discovered features around five main dimensions: i) \textit{compatibility}, the degree of compliance to standards and the availability on different operating systems, ii) \textit{image management}, the ways container engines generate, use, and share container images, iii)  \textit{optimizations}, the techniques employed to reduce resource usage, iv) \textit{container executions}, the features available at runtime, such as resource management and the support to specific libraries and tools, and v) \textit{security}, the way container engines guarantee secure image builds and executions. Note that each technology is abbreviated with the first three letters of their name (e.g., $Doc$ for Docker).

\subsection{Compatibility} 

The Open Container Initiative (OCI)~\cite{oci_website} is a Linux Foundation project to design open standards for operating-system-level virtualization, such as Linux containers. The OCI standard currently contains two specifications: the Runtime Specification (\textit{runtime-spec}) \cite{ociruntime_website} and the Image Specification (\textit{image-spec}) \cite{ociimage_website}. Given an OCI image, any container runtime that implements the OCI Runtime Specification can unbundle the image and run its contents in an isolated environment.

All selected container engines fully support the OCI standard, with the only exception of Charliecloud that only provides it as an experimental feature\footnote{ \url{https://hpc.github.io/charliecloud/command-usage.html\#ch-run-oci}}: the support is only partial and available with a set of additional commands. OCI assumes that containers be always-on services with a complex lifecycle, while Charliecloud focuses on scientific applications that are usually executed once and are terminated as soon as the required calculations are completed 
Similarly Singularity provides both OCI-compliant commands (e.g., \textit{singularity oci exec}) and additional ones (e.g., \textit{singularity exec}).

In terms of compatibility, balenaEngine is the only solution that only works on top of a dedicated operating system, namely balenaOS~\cite{balena_os_website}, that is customized and optimized for the execution of balenaEngine and containers. This Linux-based operating system, which is compatible with a wide range of computer architectures including armv5, armv6, armv7, aarch64, i386, and x86\_64, is also responsible for running a containerized daemon (device supervisor) that automatically updates running containers if an update is available.

All other engines can run on full-fledged, standard operating systems.

\vspace{0.5cm}
\begin{mdframed}
\textbf{Finding \#\findings} All analyzed container engines support OCI and standard host operating systems with a few exceptions. Docker, Podman, Sarus, and balenaEngine are fully OCI compliant. Singularity is also compatible with OCI but with a dedicated set of commands, and Charliecloud offers it as an experimental feature. balenaEngine can only run on top of its dedicated oerating system.
\end{mdframed}
\vspace{0.5cm}

\subsection{Image management} Docker can build images from a Git repository, from a compressed archive, or a Dockerfile. A Dockerfile is a text file that defines all the commands needed to build an image. 
Docker allows one to publish images onto private or public \textit{registries} that act as shared repositories, where images can be versioned through tags. This way, different versions of the same image (e.g., latest, old or beta version) can be associated with the same project.
Docker also provides a public, official registry, called \textit{Docker Hub}~\cite{docker_hub_website}, that offers a rich set of base images that can be used as a starting point for building new containerized applications. \textit{Docker Registry} allows us to create private registries. 

Podman images can be built from a Dockerfile or from a so-called \textit{Containerfile} (same syntax as Dockerfiles but with a different file extension). At its core, Podman uses Buildah~\cite{buildah_website} to build container images. Since it replicates all the commands we can find in a Dockerfile, Podman allows the user to build images with or without Dockerfiles while not requiring any root privileges. Podman supports both OCI and Docker images, and allows one to pull images from a local directory that stores all the image files, a Docker registry, or an OCI archive, that is, an image that complies with OCI. 

For image creation, Charliecloud uses external builders ---including Docker and Buildah---, or an internal one, called \textit{ch-image}. The image can be built with or without privileges, depending on the used builder. The created image is wrapped with a Charliecloud interface and converted into a compressed archive that can be easily moved on the HPC cluster. The archive is then unpacked and executed without root privileges. 

Singularity allows us to build images from external resources, as its official public registry, called \textit{Singularity Cloud Library}~\cite{singularity_cloud_library_website} (SIC), and Docker Hub, converted and used in Singularity. Singularity Definition Files (SDF), similarly to Dockerfiles, are used to create container images. An SDF is composed of two parts, the \textit{header}, which defines the configuration of the execution environment (e.g., kernel features, Linux distribution), and \textit{sections}, which execute commands during the build process. The container image can be produced in two different formats: a writable sandbox for interactive development, or in the Singularity Image Format (SIF), a compressed read-only format that can be easily moved, shared, and distributed. 

Sarus does not provide means to build images and rely on external tools. It supports already-built images that are OCI compliant, and that can be downloaded from public and private registries. 

balenaEngine uses Docker behind-the-scene for both image building and usage providing analogous capabilities. 

\vspace{0.5cm}
\begin{mdframed}
\textbf{Finding \#\findings.} All analyzed engines but Sarus provide means to build images from Dockerfiles/Containerfiles. Singularity can also use a proprietary format for both image definition and building. Non-HPC engines support images created from git repositories and archives, while all the engines allow users to share and download images using private or public registries. Charliecloud supports images loaded from archives.
\end{mdframed}
\vspace{0.5cm}

\subsection{Optimizations}

All engines provide key optimizations. All but Singularity exploit layered images, an approach pioneered by Docker.
A layered image consists of several read-only layers of data, each of them corresponding to a a single instruction (e.g., a command in a Containerfile). The layers are stacked and contain only the changes from the previous one. Layered images are particularly useful because they allow reusing any layer as starting point for a different image. Being read-only, the layers shared across different images are only stored once on the disk, and if needed, in memory. 

When a new container instance is created from an image, a writable layer, called the container layer, is also created on top of the image layers. This layer hosts all changes made to the running container, for example, it stores newly written files, modifications to existing files, and deleted files to allow for customizing the container. Changes made to the container layer do not affect image layers. This way, different images can share common files and components without the need of storing them multiple times. 

balenaEngine is built to optimize the usage of network and I/O bandwidth. When an image is updated, the devices are notified and only the differences are downloaded, and in case of faults (e.g., out of batteries) container images cannot be corrupted. When working with container images to be generated from archives, balenaEngine builds the image while uncompressing the files without the need for occupying the disk with temporary files. During the extraction, balenaEngine also reduces the memory footprint by limiting page caching.

Sarus does not provide any built-in mechanism for image creation and publication but optimizes existing images (e.g., Docker or OCI ones) by reducing their size through SquashFS~\cite{squashfs_website}, an optimization that is also employed by Singularity.

\vspace{0.5cm}
\begin{mdframed}
\textbf{Finding \#\findings.} Layered images are a widespread technique to save disk space by allowing images to share common parts. balenaEngine provides some key optimizations to reduce network transfers (i.e., differential image updates) and disk usage (i.e., on-the-fly image extraction). Sarus and Singularity compress images with SquashFS to reduce disk usage.
\end{mdframed}
\vspace{0.5cm}

\subsection{Containers executions}

When it comes to container executions, one of the main features is the support for ``detached'' executions, that is, the ability to run long-lasting interactive processes (e.g., a database management system or a web server). While applications executed in the cloud or IoT infrastructures may benefit from this feature, HPC applications are usually non-interactive and terminate as soon as the calculations are completed. For this reason, Charliecloud and Sarus do not support detached executions by design. 

While all container engines provide the means to allocate resources (e.g., cores) to a container at startup time, only Docker and Singularity allow for dynamic resource management, that is, the reconfiguration of allocated resources at runtime.

balenaEngine provides some important orchestration capabilities to deploy and update containers on a fleet of registered IoT devices, while Docker provides two dedicated tools: Docker Compose and Docker Swarm. 

\textit{Docker Compose} allows us to configure and run multi-container applications. Users write \textit{docker-compose} files to define all container images used by the application, and the instances we need of each image. Typical use cases are ``traditional'' three-tier applications or microservices where a single system is composed of multiple separated components (i.e., multiple executables). By default, Docker Compose also sets up a network for the application, and each container is reachable by the others.
\textit{Docker Swarm} lets users manage containers deployed across multiple machines. It uses the standard Docker API and networking for merging a pool of Docker hosts into a virtual, single host. It also provides means to scale applications or single containers that can be replicated onto the cluster. 
Instead, Podman relies on external tools such as Kubernetes. HPC engines do not provide ad-hoc orchestration capabilities. 

GPU computations are supported by all the engines, but Podman, assuming that the host operating system has the proper driver installed. Charliecloud provides this feature but requires non-trivial effort\footnote{\url{https://hpc.github.io/charliecloud/command-usage.html\#notes}} (e.g., recompilation) to configure some important GPU-related libraries (e.g., NVIDIA ones).  Singularity exploits GPU frameworks such as NVIDIA CUDA\footnote{\url{https://developer.nvidia.com/cuda-toolkit}} and AMD ROCm\footnote{\url{https://rocmdocs.amd.com/en/latest/}}.  In Sarus GPU devices are supported through a dedicated library, namely, the NVIDIA Container Runtime \cite{nvidiacontainer_website}  (that is OCI-compatible).

All HPC-engines support MPI libraries that help developers write complex distributed computations, and workload managers that oversee distributed executions.
Singularity supports both OpenMPI\footnote{\url{https://www.open-mpi.org}} and MPICH\footnote{\url{https://www.mpich.org}}, two popular implementations of MPI, and it provides two execution models: \textit{hybrid} and \textit{bind}. The former allows one to exploit both the MPI implementations provided by the host machine and the one (if installed) inside the containers. The latter uses only the MPI implementation available on the host. Charliecloud supports OpenMPI while Sarus only MPICH; both engines provide a dedicated set of commands and configuration files. 
As HPC workload managers, Singularity support  Torque~\cite{torque}, Slurm~\cite{slurm_website}, and SGE~\cite{sge}, while Charliecloud and Sarus can be used with Slurm.

\vspace{0.5cm}
\begin{mdframed}
\textbf{Finding \#\findings.} Detached executions are not supported by Charliecloud and Sarus because of their focus on scientific applications. Docker and Singularity provide means to dynamically allocate resources, while only Docker and balenaEngines have built-in orchestration capabilities. GPU computations are supported by all the engines except Podman. HPC-engines provide ad-hoc support for workload managers and MPI-based computations.
\end{mdframed}
\vspace{0.5cm}

\subsection{Security}

Docker is a client-server application. The server is the Docker daemon (\textit{dockerd}) that processes the requests sent by clients through a command-line interface (CLI) or a RESTful API and manages Docker images and instances. The daemon is executed with root privileges, and thus only trusted users should be allowed to control the daemon. A \textit{rootless} mode was available as an experimental feature starting from version $19.03$ and it has been fully supported since version $20.10$. The rootless mode executes the Docker daemon and containers inside an isolated user namespace (a feature that Docker itself uses also to isolate containers with one another). Thus, both the daemon and the containers run without root privileges. 

By default, Docker starts containers with a restricted set of Linux kernel capabilities \cite{kernelcap_website}\cite{linuxcap_paper} to allow us to implement a fine-grained access control system. The best practice is to remove all the capabilities except those explicitly required by the application.

Unlike Docker, Podman is daemonless and so it has no background service and the container runtime is executed only when requested. This way, Podman offers less attack surface, because it is not always in execution. It does not require root privileges by leveraging user namespaces. Both Docker and Podman require an initial configuration by a system administrator to enable rootless execution. 
Podman and Docker allows to sign images to only trust selected image providers and mitigate man-in-the-middle attacks. Docker uses a proprietary signing system, while uses  GNU Privacy Guard (GPG) before pushing the image to a remote registry. In this configuration, all the nodes running the container engine must be properly configured to retrieve the signatures from a remote server.

Similarly to Podman, Charliecloud only needs Linux user namespaces to run containers without the need of any privileged operations. It is daemonless and requires only minimal configuration changes on the computing center. It is not designed to be an isolation layer, so containers have full access to host resources. Charliecloud commands require privileged access only when used along with external builders such as Docker. 
This approach still avoids most security risks while maintaining access to the performance and functionality already offered~\cite{charliecloud_paper1} since image building is usually executed on the user machine, while only the non-privileged execution is run on in the HPC center preserving its security.

Singularity is daemonless and requires containers to have the same permissions as the users that started them, while the access to files within the container runtime is managed by the standard POSIX permissions. Containers are started with flag $PR\_NO\_NEW\_PRIVS$ that prevents applications to gain additional privileges. On multi-user, shared systems, such as HPC centers, Singularity allows an unprivileged user to run a container as root with  \textit{FakeRoot}\footnote{\url{https://wiki.debian.org/FakeRoot}}. This way, the user has almost the same administrative rights as root but only within the container. In addition, Singularity provides several strategies to ensure safety in these types of environments.
Singularity containers can be signed and verified using PGP keys and thus provide a trusted method to share containers. Singularity supports the encryption/decryption of containers at runtime to create a secure and confidential container environment. 

Sarus is also daemonless. Root permissions are required to install and work with Sarus. For this reason, Sarus checks a list of conditions to ensure that critical files and directories opened during privileged execution meet them. 

balenaEngine provides multiple security layers. Updates are sent to devices in a reliable and verifiable way to protect devices against attacks. An API key is created for each device, balenaOS then controls who can access it, the actions that are permitted, and the available communication channels. balenaEngine uses OpenVPN to control the state of its devices. The VPN disallows device-to-device traffic and prohibits outbound traffic to the Internet.
balenaEngine maintains a repository of secure base images. Any resource added to these base images is verified by a GPG key or a checksum.

\vspace{0.5cm}
\begin{mdframed}
\textbf{Finding \#\findings.} All container engines except for Charliecloud and Sarus support container image signing to increase reliability. Rootless executions are supported by Podman, Charliecloud, Singularity, and only recently by Docker. Docker and balenaEngine share the same daemon-based architecture, while the others are daemonless. balenaEngine is the only engine to provide VPN support.
\end{mdframed}

%% file: 5_evaluation.tex
\newcommand\perfeval{Performance Evaluation\xspace}
\newcommand\perfover{Overhead Evaluation\xspace}
\newcommand\perftim{startup and shutdown times}
\newcommand\perfovh{overhead}
\newcommand\perfims{image size}
\newcommand\perfmem{memory footprint}

To evaluate the performance of the six container engines we carried out an extensive empirical evaluation. We used a single-user bare-metal server equipped with an AMD CPU Ryzen 5 2600 @ 3.40GHz (6 Cores / 12 Threads), with 32GB RAM DDR4 @ 3200MHz, and Ubuntu 19.10. We exploited this configuration to avoid the performance variability introduced by hardware shared among multiple, concurrent jobs, and the overhead introduced by a virtualization system.

The tests on startup/shutdown times, \perfmem , and \perfims~used containers created from three popular images that include a ready-to-use Linux distribution: \textit{Ubuntu 18.04}, \textit{CentOS 8}, and \textit{Alpine 3}. We selected these images because they do not contain any application-level dependency that could introduce noise and bias in understanding the impact of the execution times of each engine. For the tests related to \perfovh\xspace we containerized\footnote{The source code of the containerized test suites can be found at: \url{https://github.com/deib-polimi/containers-test-suites}} the Phoronix Test Suite and we used \textit{Ubuntu 18.04} as base image since it is  the most popular Linux distribution, and a widely used solution for containers.

\subsection{Startup and shutdown times}
\label{subsec-eval_nonapp}

Since a new container instance should become available quickly, to fulfill user requests properly, and it should also die quickly, to avoid wasting resources, we measured   startuo and shutdown times. We measured them by means of the Linux command \textit{netcat}, which allows one to send data packets between a (netcat) server and a client. We installed the netcat server on the host machine and configured each container to send a packet to the host and then terminate. The startup time was measured as the difference between the time the data packet is received by the server and the instant we started the container. The shutdown time is computed as the difference between the instant the container terminated and the one the data packet was received. We also tested how the startup/shutdown times were affected when multiple containers are running concurrently on the same machine. 

\begin{table}[t]
\centering
\setlength{\tabcolsep}{2.3pt}
\begin{tabular}{llcccccc}
\toprule
\multicolumn{2}{c}{} & \multicolumn{2}{c}{\textit{Ubuntu}} & \multicolumn{2}{c}{\textit{CentOS}} & \multicolumn{2}{c}{\textit{Alpine}} \\
 & & Start & Stop & Start & Stop & Start & Stop  \\
 \midrule
\multirow{2}{*}{\textit{Doc}}  & $\mu$ &  871 & 0.1 & 889 & 0.1 & 875 & 0.1  \\
                             & $\sigma$ &  29 & 0 & 33 & 0 & 31 & 0  \\
\multirow{2}{*}{\textit{Pod}} & $\mu$ & 744 & 0.1 & 1413 & 0.1 & 559 & 0.1  \\
                             & $\sigma$ &  178 & 0 & 339 & 0 & 32 & 0  \\
\multirow{2}{*}{\textit{Sin}} & $\mu$ & 1756 & 23 & 1837 & 23 & 1740 & 23  \\
                             & $\sigma$ &  80 & 5 & 144 & 5 & 74 & 5  \\
\multirow{2}{*}{\textit{SinS}} & $\mu$ & 125 & 23 & 142 & 23 & 108 & 23 \\
                             & $\sigma$ &  8 & 6 & 10 & 5 & 7 & 5  \\
\multirow{2}{*}{\textit{Cha}} & $\mu$ & 6 & 0.1 & 7 & 0.01 & 6 & 0.1  \\
                             & $\sigma$ &  0 & 0 & 0 & 0 & 0 & 0  \\
\multirow{2}{*}{\textit{Sar}} & $\mu$ & 126 & 23 & 154 & 23 & 109 & 23  \\
                             & $\sigma$ &  9 & 5 & 11 & 5 & 7 & 5  \\
\multirow{2}{*}{\textit{bal}}  & $\mu$& 1007 & 0.1 & 1012 & 0.1 & 1010 & 0.1 \\
                             & $\sigma$ &  30 & 0 & 29 & 0 & 29 & 0  \\
\bottomrule
\end{tabular}
\caption{Startup and shutdown times in ms with single containers.}
\label{tab:sing-times}
\end{table}

\begin{figure}[thbp]
\centering
   \begin{subfigure}{0.8\textwidth}
        \includegraphics[width=\textwidth]{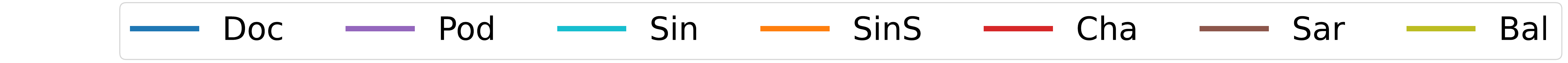}
    \end{subfigure}\\
    \vspace{0.1cm}
    \begin{subfigure}{0.8\textwidth}
        \includegraphics[width=\textwidth]{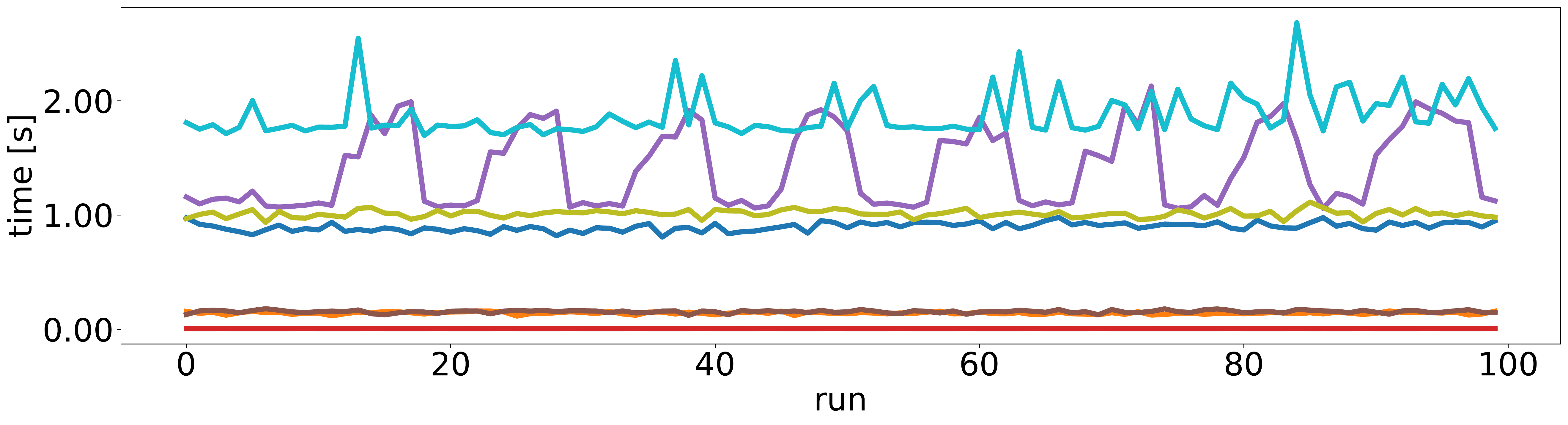}
        \caption{CentOS - Startup}
        \label{fig:single_startup_CentOS_plot}
    \end{subfigure}\\
    \begin{subfigure}{0.8\textwidth}
        \includegraphics[width=\textwidth]{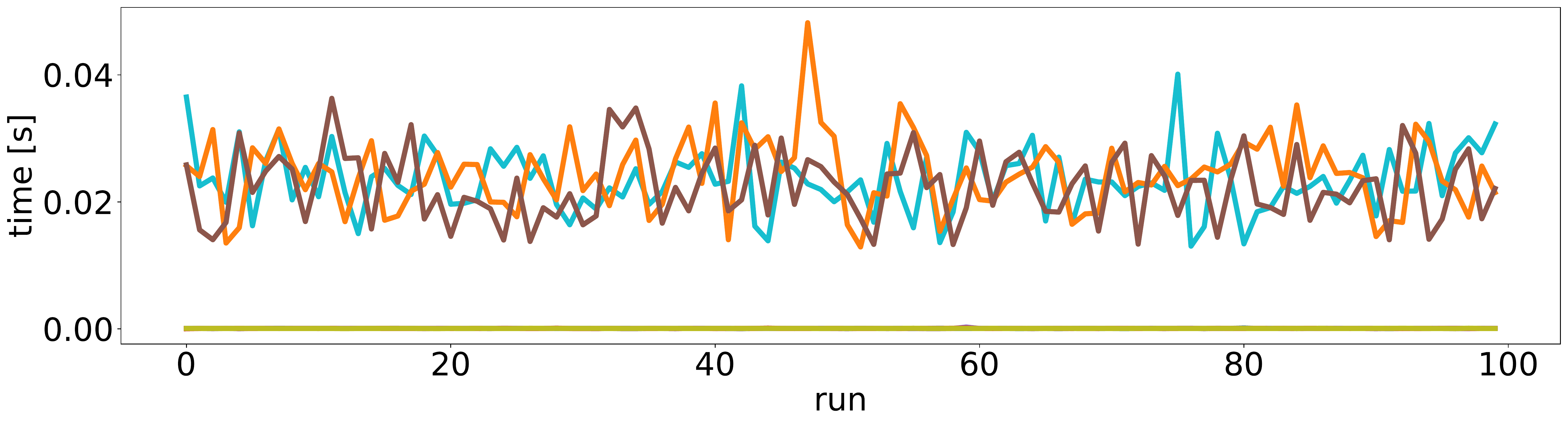}
        \caption{Ubuntu - Shutdown}
        \label{fig:single_shutdown_ubuntu_plot}
    \end{subfigure}\vspace{0.2cm}
    \caption{Example startup and shutdown times in milliseconds.}
\end{figure}

Table~\ref{tab:sing-times} shows the average ($\mu$) startup and shutdown times along with the standard deviation ($\sigma$) of each experiment executed with the three container images. Note that \textit{SinS} abbreviates Singularity with SIF images. To assess the startup and shutdown times of single containers, we repeated the measurements $100$ times, waiting 1 second before each execution and termination.

Most of the container engines exhibit a constant behaviour: startup and shutdown times remained stable during the experiment. Podman and Singularity (using Docker images) show the highest variance in startup times with multiple spikes during the experiments, especially when using the Ubuntu and CentOS images. For example, Figure~\ref{fig:single_startup_CentOS_plot} shows the startup times of the different engines running CentOS images over the $100$ repetitions.

The fastest engine in starting containers is Charliecloud, followed by Singularity with SIF images and Sarus. Docker, Podman, and balenaEngine obtained higher and quite similar startup times, while Singularity (using Docker images) is the slowest implementation. This trend is quite consistent across the three images we used. Note that even if the difference between the fastest and the slowest is some 1.7 s, the fastest is some $290$ times faster than the slowest and $20$ times faster than the second fastest engines.
Charliecloud uses a flattened and unpacked image that is ready for execution when the start command is issued and this may be the reason why it is the fastest engine. Singularity SIF and Sarus leverage SquashFS and this may help reduce the startup time compared to the others. 

Shutdown times are small, negligible, and jagged for all engines: for example, Figure~\ref{fig:single_shutdown_ubuntu_plot} shows the shutdown times of the different engines running Ubuntu images over the $100$ repetitions. We can observe that, for most engines, the startup and shutdown times do not change when using different images.
Podman and Singularity (using Docker images) are the only engines with a significant variance in startup times.

\begin{table}[htbp]
\centering
\setlength{\tabcolsep}{2.3pt}
\begin{tabular}{llcccccc}
\toprule
\multicolumn{2}{c}{} & \multicolumn{2}{c}{\textit{Ubuntu}} & \multicolumn{2}{c}{\textit{CentOS}} & \multicolumn{2}{c}{\textit{Alpine}} \\
 & & Start & Stop & Start & Stop & Start & Stop  \\
 \midrule
\multirow{2}{*}{\textit{Doc}}  & $\mu$ &  631 & 402 & 624 & 395 & 610 & 10435  \\
                             & $\sigma$ &  45 & 33 & 53 & 31 & 45 & 31  \\
\multirow{2}{*}{\textit{Pod}} & $\mu$ & 1645 & 648 & 4988 & 1115 & 474 & 10329  \\
                             & $\sigma$ & 1055 & 533 & 1894 & 651 & 213 & 35  \\
\multirow{2}{*}{\textit{Sin}} & $\mu$ & 2102 & 78 & 1912 & 82 & 1919 & 78  \\
                             & $\sigma$ &  80 & 10 & 75 & 11 & 73 & 10  \\
\multirow{2}{*}{\textit{SinS}} & $\mu$ & 304 & 78 & 306 & 82 & 302 & 77 \\
                             & $\sigma$ &  7 & 10 & 11 & 11 & 9 & 9  \\
\multirow{2}{*}{\textit{bal}}  & $\mu$& 748 & 456 & 753 & 455 & 747 & 10501 \\
                             & $\sigma$ &  60 & 45 & 70 & 42 & 76 & 37  \\
\bottomrule
\end{tabular}
\caption{Startup and shutdown times in ms with multiple containers.}
\label{tab:mul-times}
\end{table}


\begin{figure}[htbp]
\centering
  \begin{subfigure}{0.8\textwidth}
    \includegraphics[width=\textwidth]{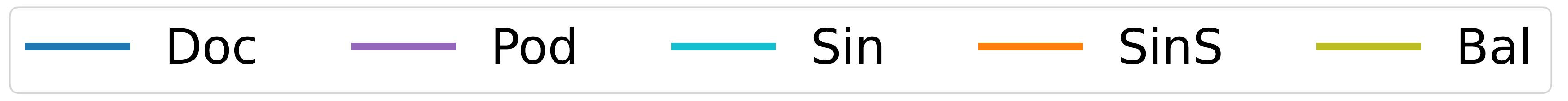}
    \end{subfigure} \\
    \begin{subfigure}{0.8\textwidth}
        \includegraphics[width=\textwidth]{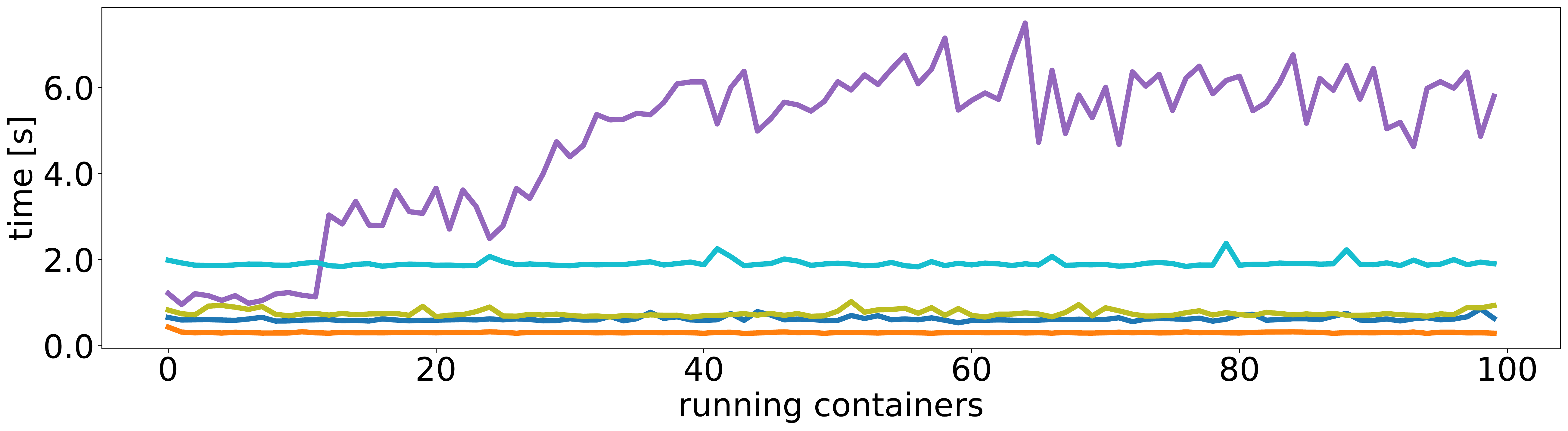}
        \caption{CentOS - Startup}
        \label{fig:sconcurrent_startup_CentOS_plot}
    \end{subfigure}
    \begin{subfigure}{0.8\textwidth}
        \includegraphics[width=\textwidth]{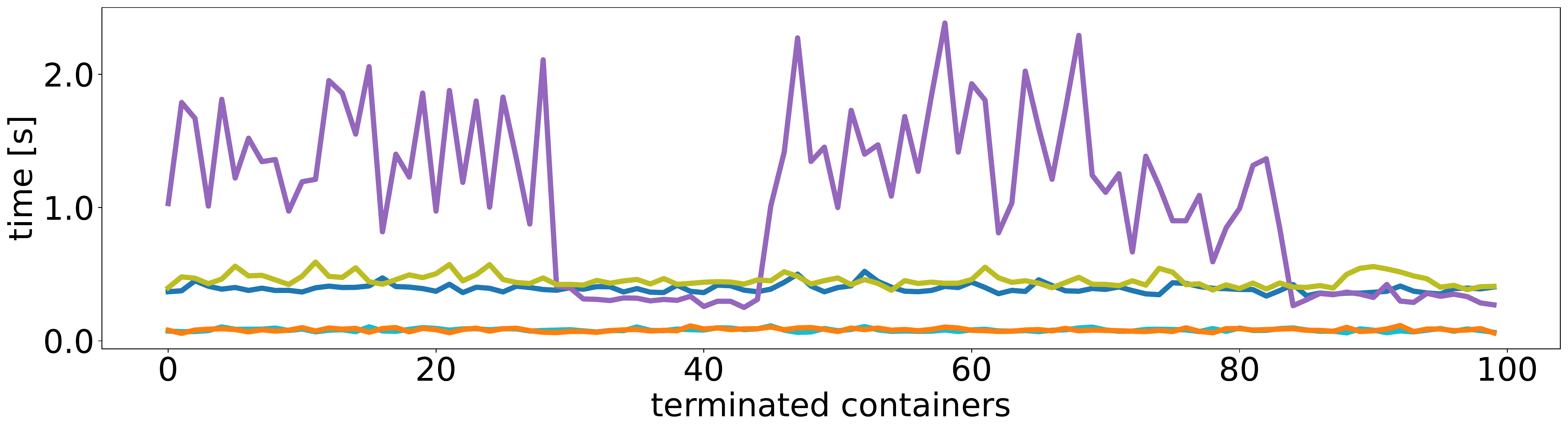}
        \caption{CentOS - Shutdown}
        \label{fig:sconcurrent_shutdown_CentOS_plot}
    \end{subfigure}\vspace{0.2cm}
    \caption{Example concurrent startup and shutdown times in milliseconds.}
    \label{fig:concurrent_startup_times_plot}
\end{figure}

We also measured the startup and shutdown times when multiple containers are executed concurrently on the same machine. We started with a container and created a new one as soon as the previous one was up and running, incrementally up to $100$ concurrent instances. We do not report Charliecloud and Sarus because they do not provide the means to run containers in detached mode (i.e., in parallel in the background). 

Table~\ref{tab:mul-times} shows the average startup and shutdown times along with the standard deviations. The times obtained by Docker, Singularity (with and without SIF), and balenaEngine are almost constant while the number of running containers increases. On the other hand, Podman shows a noticeable delay during the experiment with the Ubuntu and CentOS images. For example, Figure~\ref{fig:sconcurrent_startup_CentOS_plot} shows
the execution of the experiment with CentOS images and we can observe that the startup time increases at around the execution of the 12th container and again at around the launch of the 22nd. Podman also shows some spikes when running the Alpine image. 

Podman exhibits a strong variability during the shutdown phase, especially with Ubuntu and CentOS images (Figure~\ref{fig:sconcurrent_shutdown_CentOS_plot} refers to the experiment with CentOS images). Other implementations present a constant behavior and Singularity (using both Docker and SIF images) is significantly the fastest engine in stopping containers. Shutdown times obtained by the other technologies are particularly high (around 10 seconds) when using a small image such as Alpine while they are small when using Ubuntu or CentOS. The variance is negligible for all the engines but Podman.
 
\vspace{0.5cm}
\begin{mdframed}
\textbf{Finding \#\findings.} HPC-optimized container engines  (Charliecloud, Singularity, and Sarus) are the fastest in terms of startup times. Charliecloud obtained results that are at least two orders of magnitude better than the other implementations. Singularity is fast but only when used with SIF images. Concurrent images do not affect the startup times except for Podman that showed a significant variance. Shutdown times are usually negligible and almost constant but concurrent executions may slow down the operation.
\end{mdframed}

\subsection{Memory footprint}
\label{subsec-perfmem}

We measured the memory footprint to access the memory consumed by an engine to instantiate and execute images. The containerized executable should consume the same amount of memory as the non-containerized version. The memory should only be freed when the container is stopped and deleted to make room for other containers. We obtained it by means of the Linux command \textit{free -m}, which provides information about the total amount of physical and swap memory, and with an engine-specific command if available. We also measured the evolution of memory allocation during container creation using command \textit{nmond}, which supplies a benchmark tool to collect performance data regarding memory and other resources\footnote{\url{https://www.ibm.com/docs/en/aix/7.2?topic=n-nmon-command}.}.

\begin{table}[t]
\centering
\begin{tabular}{lccc}
\toprule
 & Ubuntu & CentOS  & Alpine  \\
   \midrule
\textit{Doc}& 28 MB & 27 MB & 27 MB \\
\textit{Pod}& 15 MB & 14 MB & 17 MB \\
\textit{Sin} & 4 MB & 3 MB & 2 MB \\
\textit{SinS}  & 4 MB & 3 MB & 2 MB \\
\textit{Cha}& 0 MB & 0 MB & 0 MB \\
\textit{Sar} & 4 MB & 6 MB & 4 MB \\
\textit{bal} & 26 MB & 28 MB & 26 MB \\
\bottomrule
\end{tabular}
\caption{Memory footprints.}
\label{tab:mem_comparison}
\end{table}

Table~\ref{tab:mem_comparison} shows the memory footprints we measured ---using command \textit{free}--- of idle container instances (without any user interaction). The memory allocated by Charliecloud is zero: no memory is allocated to the container when it is idle. Singularity (both cases) showed a small memory footprint ($2-4$MB), similarly to Sarus ($4-6$MB). Podman provided a significantly higher result ($14-17$MB), but Docker ($27-28$MB) and balenaEngine ($26-28$MB) are the engines with the highest memory footprint and obtained similar results being based on the same underlying foundation.

\begin{figure}[t]
    \centering
    \begin{subfigure}[b]{1\textwidth}
        \includegraphics[width=\textwidth]{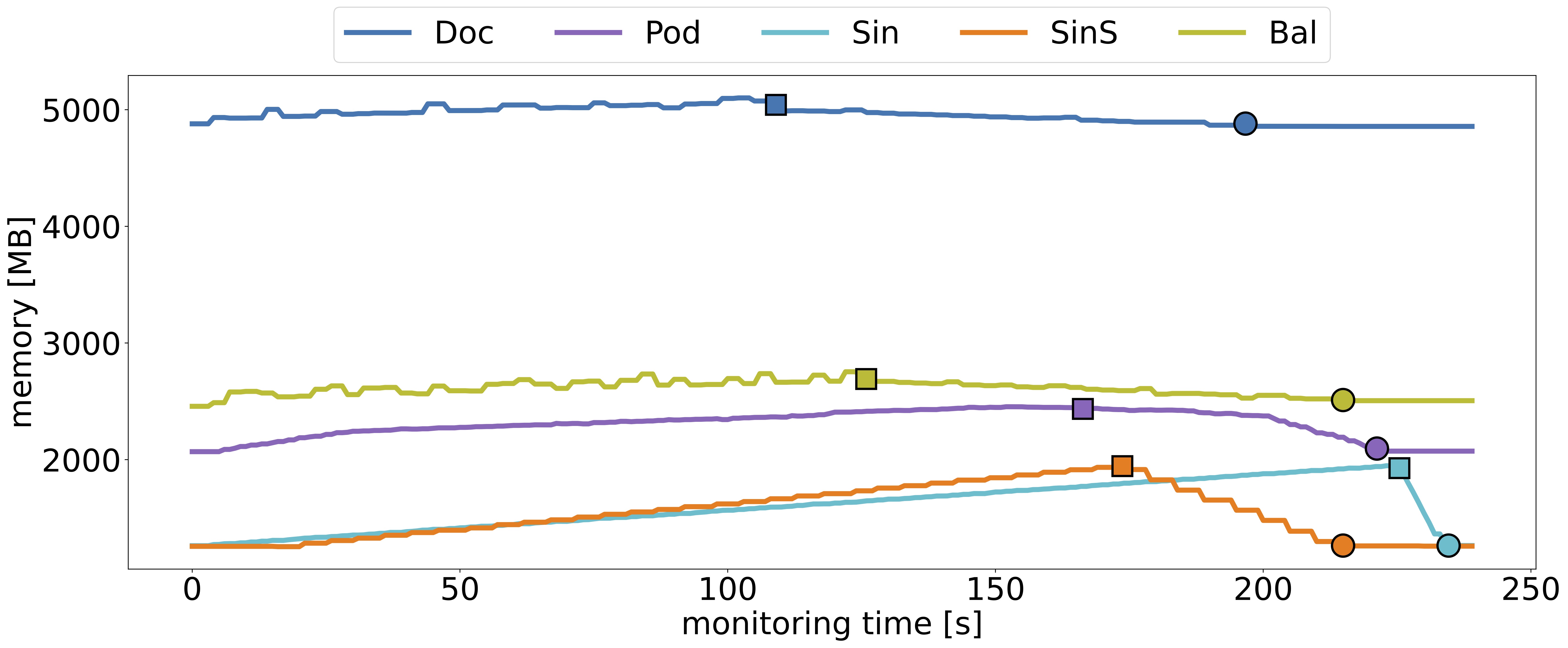} 
        \caption{Evolution (Ubuntu image). Squares and circles highlight the end of  container creation and termination, respectively.}
        \label{fig:active_memory_mon_ubuntu}
    \end{subfigure}
    \begin{subfigure}[b]{0.5\textwidth}
\includegraphics[width=\textwidth]{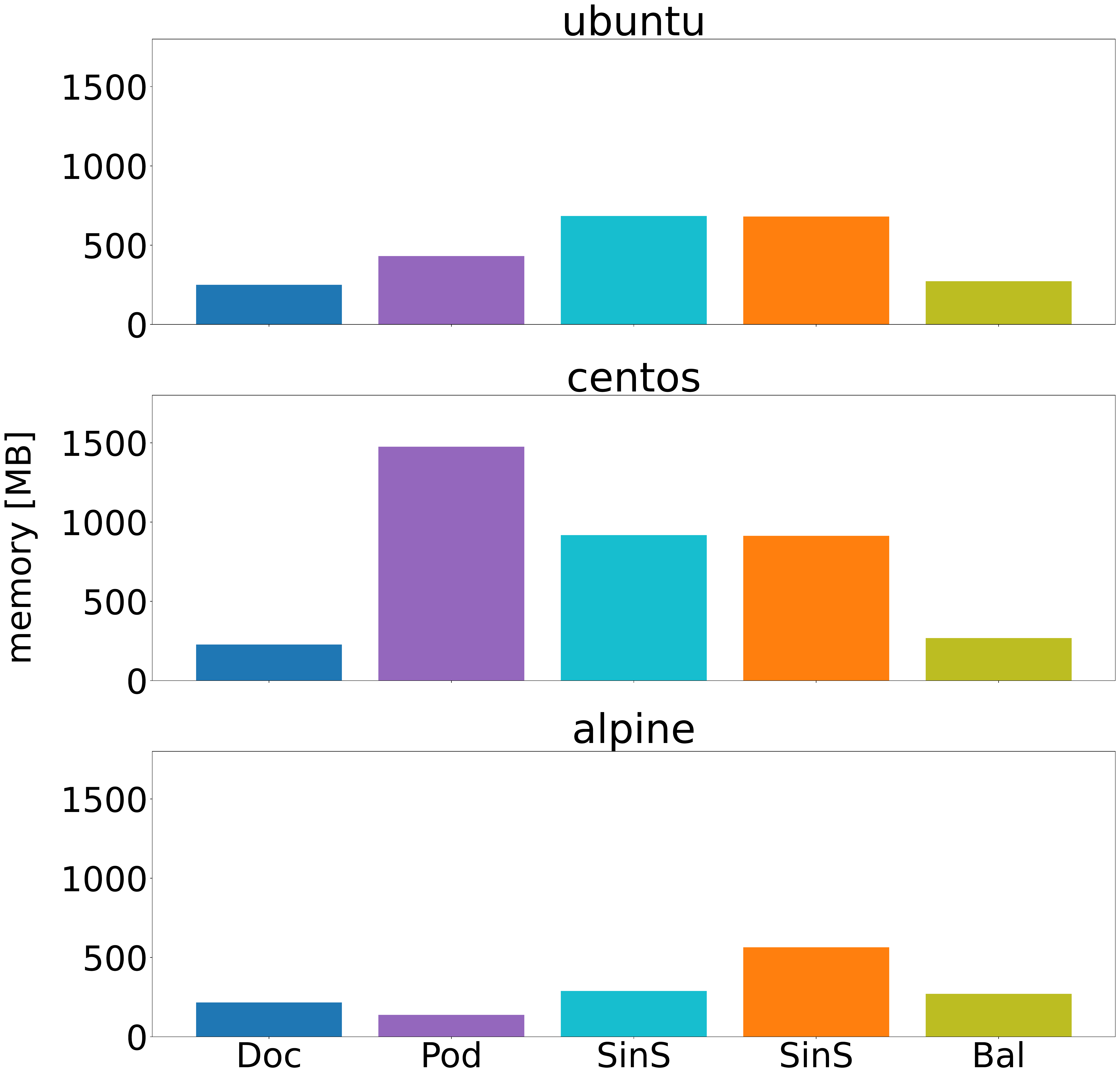}
        \caption{Comparison}
        \label{fig:active_memory_bar}
    \end{subfigure}
    \caption{Memory allocation during container creation.}
    \label{fig:active_memory}
\end{figure}

Figure \ref{fig:active_memory_mon_ubuntu} materializes the evolution of the allocated memory during the creation of $100$ container instances from image Ubuntu. As in the previous experiment, one instance was added as soon the creation of the previous one was concluded, up to $100$ containers. When $100$ instances were created, we started terminating them one after the other with the same rationale used for their creation.  Given that Charliecloud and Sarus do not allow one to run containers in the background, we did not consider them for this experiment.
 
In the creation phase, the memory grows linearly. We observed some spikes with Docker and balenaEngine, while Singularity and Podman are more stable.  During the termination phase, Singularity (without SIF) appeared to deallocate the memory quite fastly with an almost vertical descent; the other engines, instead, do it more gradually. In this experiment, Docker is the fastest to complete the creation phase and the overall process. 
 Figure \ref{fig:active_memory_bar} shows the delta of the memory allocated (from the beginning of the experiment to the creation of the 100th container) by the different engines for the three images. Results appear to depend on the considered image. Singularity allocates more memory than the others with Ubuntu and Alpine, while Podman with CentOS. We did not notice any noticeable memory leakage during the experiments. 

\vspace{0.5cm}
\begin{mdframed}
\textbf{Finding \#\findings.} Charliecloud appears to use memory efficiently only if the container is not idle. Singularity and Sarus have a significant smaller memory footprint compared to Podman, while Docker and balenaEngine require  much more memory. When multiple concurrent container creations/terminations are executed Docker is overall the more efficient in completing the whole experiments, while results for other engines appear to be image-dependent. 
\end{mdframed}
\vspace{0.5cm}

\subsection{Image sizes}
\label{subsec-perfims}

We focused on the image size of containerized applications since it should be proportional to the packaged application along with its dependencies. The ability to reuse data (e.g., libraries) contained in other images and image compression techniques can help save disk space. We measured it in three ways: i) through the data reported in the registry that the engines use, ii) with the Linux command \textit{du}, which allows a user to gain disk usage information, to obtain the size of the image on the disk, and iii) with an engine-specific command if available. For Singularity, we run the tests both with Docker images and SIF images. 

\begin{table}[t]
\centering
\begin{tabular}{l|cc|cc|cc}
\toprule
 & \multicolumn{2}{c|}{\textbf{Ubuntu}} & \multicolumn{2}{c|}{\textbf{CentOS}} & \multicolumn{2}{c}{\textbf{Alpine}} \\
 & \textit{Registry} & \textit{Disk} & \textit{Registry} & \textit{Disk} & \textit{Registry} & \textit{Disk} \\
    \midrule
\textit{Doc}& 25 MB\textsuperscript{a} & \begin{tabular}[c]{@{}l@{}}64 MB\textsuperscript{c}\\ 64 MB\textsuperscript{i}\end{tabular} & 70 MB\textsuperscript{a} & \begin{tabular}[c]{@{}l@{}}237 MB\textsuperscript{c}\\ 234 MB\textsuperscript{i}\end{tabular} & 3 MB & \begin{tabular}[c]{@{}l@{}}6 MB\textsuperscript{c}\\ 6 MB\textsuperscript{i}\end{tabular} \\\midrule 
\textit{Pod}& 25 MB\textsuperscript{a} & \begin{tabular}[c]{@{}l@{}}67 MB\textsuperscript{d}\\ 64 MB\textsuperscript{i}\end{tabular} & 70 MB\textsuperscript{a} & \begin{tabular}[c]{@{}l@{}}245 MB\textsuperscript{d}\\ 234 MB\textsuperscript{i}\end{tabular} & 3 MB\textsuperscript{a} & \begin{tabular}[c]{@{}l@{}}6 MB\textsuperscript{d}\\ 6 MB\textsuperscript{i}\end{tabular} \\\midrule 
\textit{Sin} & 53 MB\textsuperscript{b} & \begin{tabular}[c]{@{}l@{}}109 MB\textsuperscript{g}\\ 53 MB\textsuperscript{h}\end{tabular} & 80 MB\textsuperscript{b} & \begin{tabular}[c]{@{}l@{}}293 MB\textsuperscript{g}\\ 81 MB\textsuperscript{h}\end{tabular} & 3 MB\textsuperscript{b} & \begin{tabular}[c]{@{}l@{}}6 MB\textsuperscript{g}\\ 3 MB\textsuperscript{h}\end{tabular} \\\midrule 
\textit{Cha}& 25 MB\textsuperscript{a} & \begin{tabular}[c]{@{}l@{}}70 MB\textsuperscript{g}\\ 25 MB\textsuperscript{i}\end{tabular} & 70 MB\textsuperscript{a} & \begin{tabular}[c]{@{}l@{}}252 MB\textsuperscript{g}\\ 69 MB\textsuperscript{i}\end{tabular} & 3 MB\textsuperscript{a} & \begin{tabular}[c]{@{}l@{}}6 MB\textsuperscript{g}\\ 3 MB\textsuperscript{i}\end{tabular} \\\midrule 
\textit{Sar} & 25 MB\textsuperscript{a} & 25 MB\textsuperscript{e} & 70 MB\textsuperscript{a} & 67 MB\textsuperscript{e} & 3 MB\textsuperscript{a} & 3 MB\textsuperscript{e}
\\\midrule 
\textit{bal} & 25 MB\textsuperscript{a} & \begin{tabular}[c]{@{}l@{}}64 MB\textsuperscript{f}\\ 64 MB\textsuperscript{i}\end{tabular} & 70 MB\textsuperscript{a} & \begin{tabular}[c]{@{}l@{}}237 MB\textsuperscript{f}\\ 234 MB\textsuperscript{i}\end{tabular} & 3 MB & \begin{tabular}[c]{@{}l@{}}6 MB\textsuperscript{f}\\ 6 MB\textsuperscript{i}\end{tabular} \\
\bottomrule
\end{tabular}
\caption{Image sizes. Data retrieved from [a] \textit{DockerHub}, [b] \textit{Singularity Library}, [c] \textit{docker images}, [d] \textit{podman images}, [e] \textit{sarus images}, [f] \textit{balena images}, [g] folder size, [h] SIF size, and [i] tar file size.}
\label{tab:imagesize_comparison}
\end{table}

Table \ref{tab:imagesize_comparison} shows the sizes of used images retrieved from a public registry (compressed size, calculated by summing up the size of each image layer) and from the disk (compressed and uncompressed sizes). Note that all the engines, except Singularity that leverages Singularity Hub, use Docker Hub as default public registry.
We observed that the size on disk can often exceed the one reported on the registry. Registry data may not consider all the layers of the image and report the compressed size. 

Docker, Podman, and balenaEngine allow one to save the image as an archive, and the image size does not change (significantly) when exported. Singularity exploits SIF and SquashFS to compress images, which results in an important reduction of the image size on disk. Charliecloud and Sarus require the smallest images. Charliecloud provides means to store images as compressed archives (i.e., tar files) while Sarus uses SquashFS to optimize image sizes. 

\vspace{0.5cm}
\begin{mdframed}
\textbf{Finding \#\findings.} Docker, Podman, and balenaEngine do not provide any significant optimization to save disk space. SIF helps Singularity reduce significantly the size on disk (around three times smaller than Docker ones in the best case). Charliecloud and Sarus are the best engines in terms of image size (around 50\% smaller images than Singularity in the best case).
\end{mdframed}

\subsection{Overhead}
\label{subsec-perfovh}

Container engines introduce a computational overhead to run images. Ideally, the engine should provide performance comparable to the one obtained on bare-metal machines. We measured it by using the Phoronix Test Suite (PTS)~\cite{phoronix_website}, which allows one to run open-source benchmarks fetched from \textit{OpenBenchmarking.org}~\cite{openbenchmarking_website} and provides means to combine them in test suites. In particular, we defined a \textit{resource-level} suite and an \textit{app-level} one. The former performs some low-level tests to measure CPU, memory, disk, and network consumption and includes the following benchmarks\footnote{More information about each benchmark can be found at: \url{https://openbenchmarking.org/tests}}: \textit{OSBench},  \textit{PyBench},  \textit{Threaded I/O Tester},  \textit{RAMspeed SMP},  \textit{7-Zip Compression},  \textit{OpenSSL}, and  \textit{Sockperf}. The latter executes tests at application level and includes benchmarks that use Apache HTTP Server, nginx Server, BlogBench, SQLite, Redis, and Apache Cassandra.

\begin{table*}[t]
\footnotesize
\centering
\setlength{\tabcolsep}{2pt}
\begin{tabular}{lcccccccccccccccc}
\toprule
 & \multicolumn{5}{c}{\textbf{OSBench}} & \textbf{Py} & \textbf{7-Zip }& \textbf{SSL} & \multicolumn{3}{c}{\textbf{RamSpeed}} & \multicolumn{2}{c}{\textbf{I/O}} & \multicolumn{3}{c}{\textbf{sockperf}} \\
 & F & P & T & P & MA &  &  &  & Add & Avg & Copy & RE & WR & TP & PPL & LL \\
 \midrule
\textit{Doc}& -41\% & -7\% & -4\% & -4\% & 0\% & -29\% & -2\% & 0\% & 0\% & 0\% & 0\% & -7\% & 0\% & 25\% & 5\% & -7\% \\
\textit{Pod}& -4\% & -5\% & -7\% & 0\% & -5\% & -29\% & -3\% & 0\% & 1\% & 1\% & 0\% & -9\% & 0\% & 22\% & 8\% & 7\% \\
\textit{Sin} & 1\% & -5\% & 3\% & -5\% & 0\% & 0\% & 1\% & 0\% & 0\% & 0\% & 0\% & -2\% & 1\% & 6\% & 1\% & 5\% \\
\textit{SinS}  & 2\% & -1\% & 5\% & -3\% & 3\% & 0\% & 0\% & 0\% & 0\% & 0\% & 0\% & -4\% & -3\% & 4\% & 2\% & 2\% \\
\textit{Cha}& 4\% & 9\% & 15\% & 5\% & 1\% & 0\% & -1\% & 0\% & 0\% & 0\% & 1\% & -17\% & -3\% & 2\% & 0\% & 3\% \\
\textit{Sar} & 4\% & -1\% & 4\% & 0\% & 1\% & 0\% & 1\% & 0\% & 0\% & 0\% & 0\% & -2\% & -4\% & 2\% & -5\% & 7\% \\
\textit{bal} & -12\% & -6\% & -2\% & -1\% & 0\% & -29\% & -2\% & 0\% & 0\% & 0\% & 0\% & -10\% & -4\% & 23\% & 3\% & -5\% \\
\bottomrule
\end{tabular}
\caption{\textit{resource-level} test suite.}
\label{tab:hardware_suite}
\end{table*}

Tables~\ref{tab:hardware_suite} and~\ref{tab:cloud_suite} show the overheads measured with the two test suites: resource- and application-level, respectively. Each value in the tables refers to the percentage difference between container-based and bare-metal executions: a negative value means that containers perform worse than bare metal. Each test was repeated 5 times and we present the average values. 

The \textit{resource-level} test suite shows i) the amount of files created ($F$), processes ($P$), threads ($T$), and programs ($P$) started, and memory allocated ($MA$) through OSBench; ii) the time taken to complete the execution of the Python programs in PyBench ($Py$); iii) the time needed to archive/unarchive the files contained in 7-Zip; iv) the speed ($sign/s$) in generating cryptographic signatures with OpenSSL ($SSL$); v) the performance of memory operations \textit{add} and \textit{copy}, and the average ($Avg$) with RamSpeed;  vi) the bytes read ($RE$) and written ($WR$) on disk during Threaded I/O Tester; and vii) the throughput ($TP$), ping-pong latency ($PPL$), and load latency ($LL$) measured with sockperf.

As for the values obtained with OSBench, most container engines perform similar to bare-metal, except for file generation with Docker ($-41\%$), and thread management with Charliecloud ($+15\%$). Singularity, Charliecloud, and Sarus obtained results equal to bare-metal ones with PyBench, while the other engines performed significantly worse ($-29\%$). We measured no overheads with 7-Zip, OpenSSL, and RamSpeed for all the engines. The reading performance obtained with test I/O appears to be worse than bare-metal for Charliecloud ($-17\%$), balenaEngine ($-10\%$), Podman ($-9\%$) and Docker ($-7\%$). The majority of Sockperf results obtained with containers are better than bare metal with a significant gain in throughput with Docker, Podman, and balenaEngine. \blue{Most of the reported data are very close to the ones of bare metal indicating that containers add a negligible overhead to computations. Being these tests executed on the cloud, both small positive and negative variations may be introduced by  disturbances (e.g., resource contention) at the underlying hardware infrastructure. }

\begin{table}[t]
\centering
\setlength{\tabcolsep}{5pt}
\begin{tabular}{lccccccccccc}
\toprule
 & \textbf{Apache} & \textbf{nginx} & \multicolumn{2}{c}{\textbf{BlogBench}} & \multicolumn{2}{c}{\textbf{SQLite}} & \multicolumn{2}{c}{\textbf{Redis}} & \textbf{Cassandra} \\
 &  &  & Read & Write & 1T & 32T & Get & Set &  \\
  \midrule
\textit{Doc}& 15\% & -16\% & -14\% & -6\% & -1\% & 0\% & 0\% & 0\% & -2\% \\
\textit{Pod}& 19\% & 22\% & -1\% & -1\% & 0\% & 0\% & 5\% & 0\% & 1\% \\
\textit{Sin} & 0\% & 0\% & -1\% & 0\% & 0\% & -1\% & 3\% & 4\% & 0\% \\
\textit{SinS}  & 0\% & 0\% & 0\% & 0\% & 0\% & 1\% & 3\% & -1\% & 7\% \\
\bottomrule
\end{tabular}
\caption{\textit{app-level} test suite.}
\label{tab:cloud_suite}
\end{table}

The results with the \textit{app-level} test suite refer to Apache Web server, nginx, BlogBench, which measures the performance in reading and writing blog posts, SQLite, configured with both 1 and 32 threads, Redis, and Cassandra. Note that these applications are mostly used in cloud-based deployments since they rely on always-on processes, which are not well suited for HPC enviroments, and they are heavily stateful and do not fit the typical ``ephemeral'' and stateless IoT computations. For these reasons, we do not report data for Charliecloud, Sarus and balenaEngine. \blue{We report the results for Singularity since it is sometimes used as a general purpose container engine (e.g., for running web applications\footnote{https://sylabs.io/2018/09/nodejs-on-singularity/})}. 

Table \ref{tab:cloud_suite} shows that container engines demonstrated performance similar to or better than bare-metal with all applications, except for Docker with nginx ($-16\%$) and with BlogBench-Read ($-14\%$). \blue{These tests confirm the results reported for \textit{resource-level} experiments: in general containers do not introduce a significant overhead to computation even when tested with well-known middleware software. Singularity and Podman provide performance that are very close to bare-metal, while we observed small performance degradation with Docker. }

\vspace{0.5cm}
\begin{mdframed}
\textbf{Finding \#\findings.} Overall, the performance  obtained by container engines are very similar to the ones measured for bare-metal executions. Docker reported a significant overhead in a few experiments (OSBench file generation, nginx, and blog post read), while all the non-HPC engines showed around 30\% worse performance compared to bare-metal in Python-related tests. \blue{We observed that HPC-engines provide a smaller overhead compared to other engines, with performance that are almost identical to bare-metal. The maximum difference in performance compared to bare-matel is 5\% for Sarus and Singularity, 17\% for Charliecloud, 29\% for Podman and balenaEngine, and 41\% for Docker.}
\end{mdframed}

%% file: 6_hpc.tex
The last set of experiments is dedicated to evaluating the performance of HPC-specific container engines. In particular, the goal of the tests is to analyze the overhead introduced during the execution of computationally intensive and highly parallel applications. 

\subsubsection{Experiment Setup}
\label{hpc:cluster_setup}

We deployed a cluster of high-performance virtual machines on Microsoft Azure. We used machines of type HB60rs equipped with 60 cores and 240GB of RAM. These machines also feature a network interface for RDMA (remote direct memory access) connectivity that allows processes to communicate over an InfiniBand network (100 Gb/sec Mellanox EDR with single root input/output virtualization). For the experiments, we used up to 8 instances of HR60rs VMs.

We initially used Azure Cyclecloud, a tool provided by Microsoft, for the configuration of HPC clusters.
Since the tool was quite unstable and slow, we created a set of Ansible playbooks~\footnote{\url{https://github.com/deib-polimi/containers-SlurmCluster}} to automate the management operations. We set up the cluster with the Slurm workload manager and a distributed file system (NFS\footnote{\url{http://nfs.sourceforge.net}}). We used the Azure CentOS VM image since it was designed for HPC applications and embeds all the necessary drivers to work with InfiniBand and MPI. We used OpenMPI 4, configured in \textit{hybrid} mode so that both the library installed in the host node and the one in the container can be used (see Section~\ref{sec-features}).

We built the container images with Docker to use the same solution across the different engines. We created a Dockerfile\footnote{\url{https://github.com/deib-polimi/containers-DockerHPC}}, starting from the one provided by Azure\footnote{\url{https://github.com/Azure/batch-shipyard/tree/master/recipes/mpiBench-Infiniband-OpenMPI}} that includes support for InfiniBand and OpenMPI. In addition, we updated the drivers to match the one installed on the host VM and we included the benchmarks we wanted to execute.
We used two well-known test suites for MPI: the OSU Micro-Benchmarks\footnote{\url{http://mvapich.cse.ohio-state.edu/benchmarks/}}  by the Ohio State University, and mpiBench\footnote{\url{https://github.com/LLNL/mpiBench}} by the Lawrence Livermore National Laboratory. 

While we successfully configured both Singularity and Charliecloud, we could not make Sarus properly execute the benchmarks by using the envisaged OpenMPI tools. However, the team behind Sarus executed some of the benchmarks we used on its own, and the results are publicly available~\cite{sarusosu}.

We run three types of experiments: \textit{latency}, \textit{all-to-all}, and \textit{scatter/gather}. The first two are part of the OSU benchmark, while the third is included in mpiBench. Test \textit{latency} measures the min, max, and average latency of a ping-pong communication between two processes. The messages are sent repeatedly with different payload sizes to measure one-way latency. Test \textit{all-to-all} measures the min, max, and average latency of the $MPI\_Alltoall$ operation in which each  node sends (receives) a message to (from) all the other nodes, among \textit{all} involved processes.  Finally, test \textit{scatter/gather} measures the latency of sending messages (between 0 and 64kB) through the MPI primitives \textit{scatter} and \textit{gather}. 
To automate the execution of these benchmarks with Slurm, we developed a set of scripts\footnote{\url{https://github.com/deib-polimi/containers-MPIBenchmarksBatch}}. The container images were fetched from local disks to avoid any performance degradation due to network overhead.

\subsubsection{Results}
\label{hpc:results_osu}

\begin{figure}[p]
\centering
    \begin{subfigure}{0.49\textwidth}
        \includegraphics[width=\textwidth, trim={0 0 0 2cm},clip]{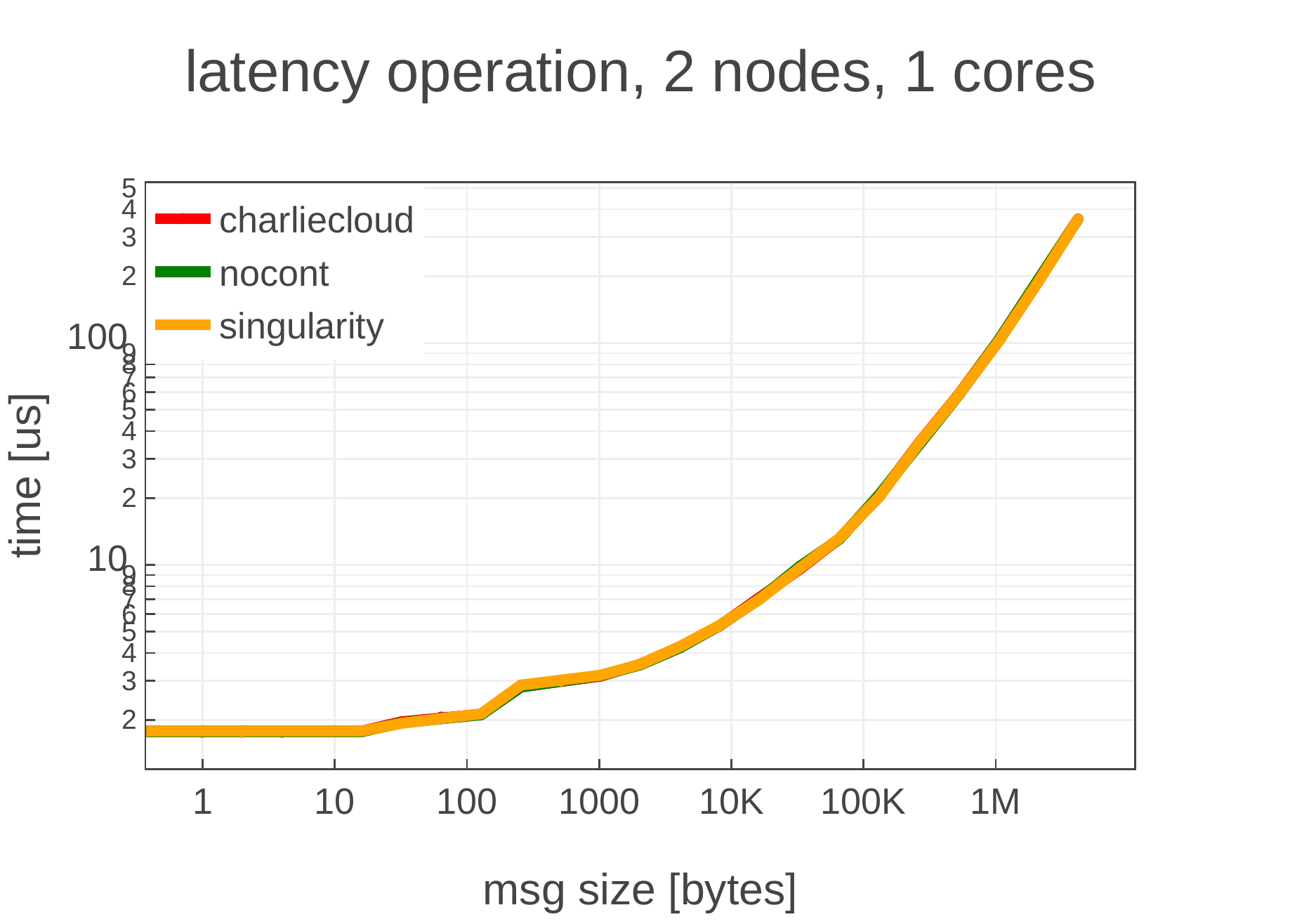} 
        \caption{Test \textit{latency} - 2 nodes - 1 core}
        \label{fig:osu_latency_2nodes_1c}
    \end{subfigure}
    \begin{subfigure}{0.49\textwidth}
        \includegraphics[width=\textwidth, trim={0 0 0 2cm},clip]{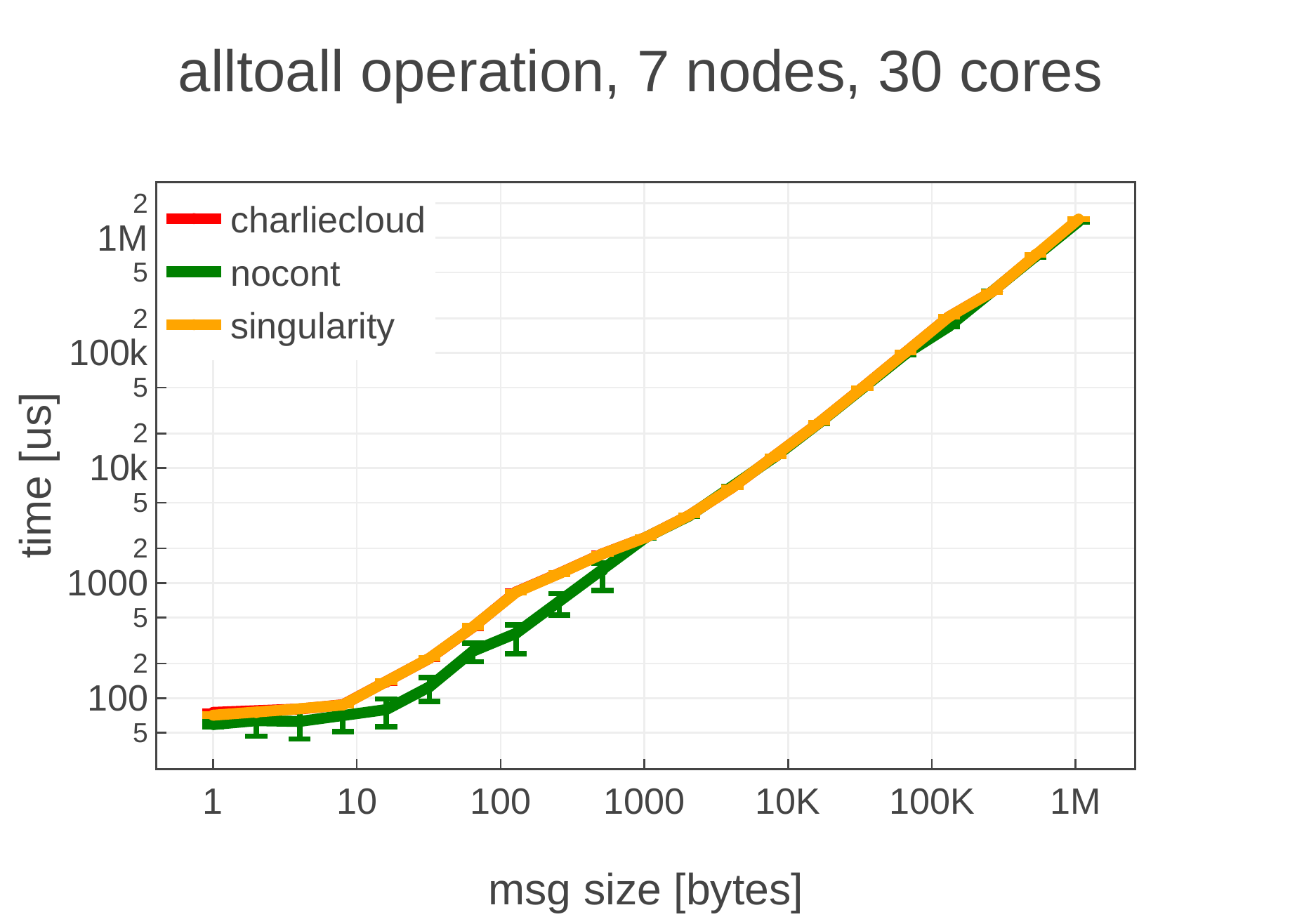}
        \caption{Test \textit{all-to-all} - 7 nodes}
        \label{fig:osu_alltoall_7nodes_30c}
    \end{subfigure}
    \begin{subfigure}{0.49\textwidth}
        \includegraphics[width=\textwidth, trim={0 0 0 2cm},clip]{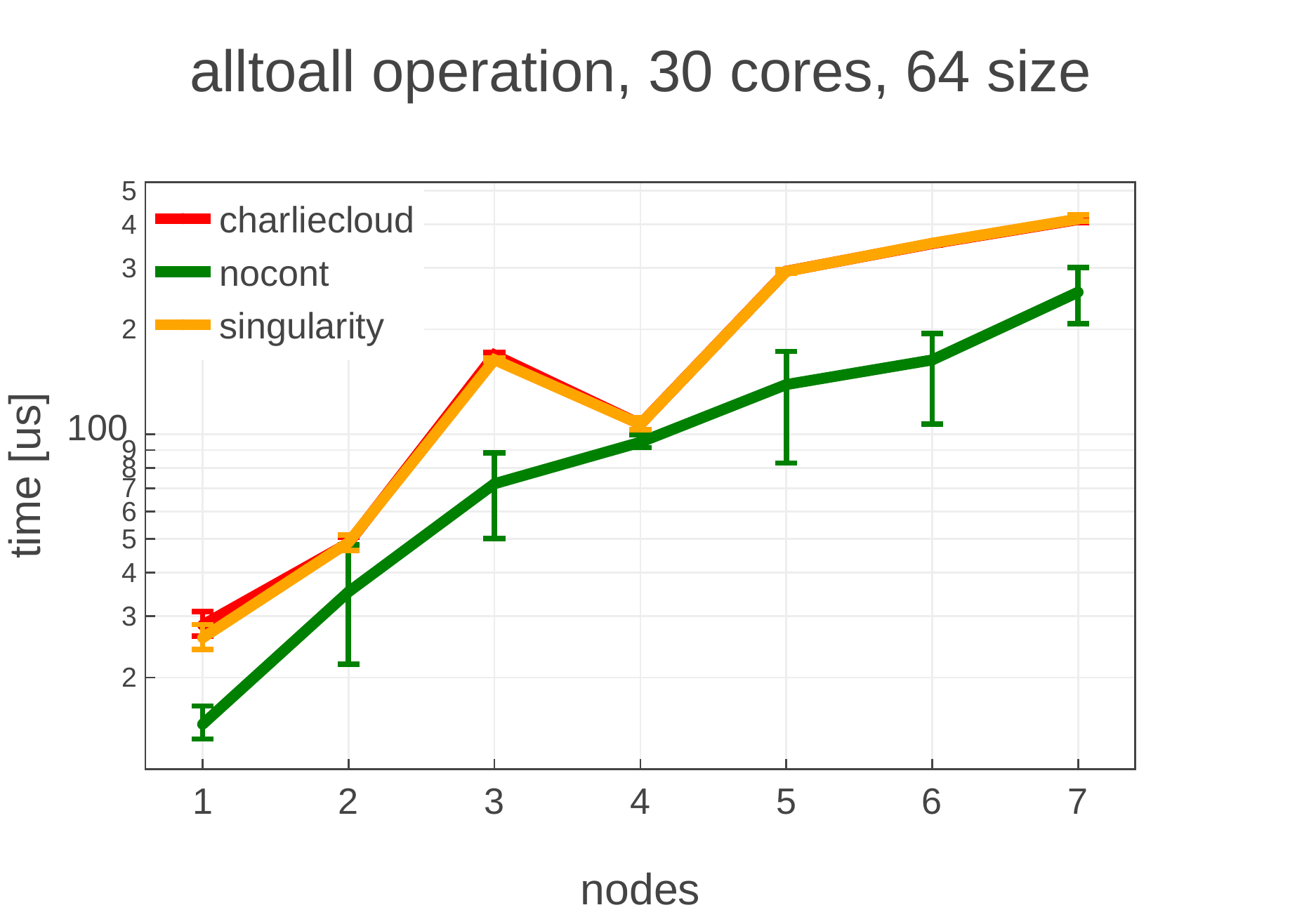} 
        \caption{Test \textit{all-to-all} - 64 bytes}
        \label{fig:osu_alltoall_64_30c}
    \end{subfigure}
    \begin{subfigure}{0.49\textwidth}
        \includegraphics[width=\textwidth, trim={0 0 0 2cm},clip]{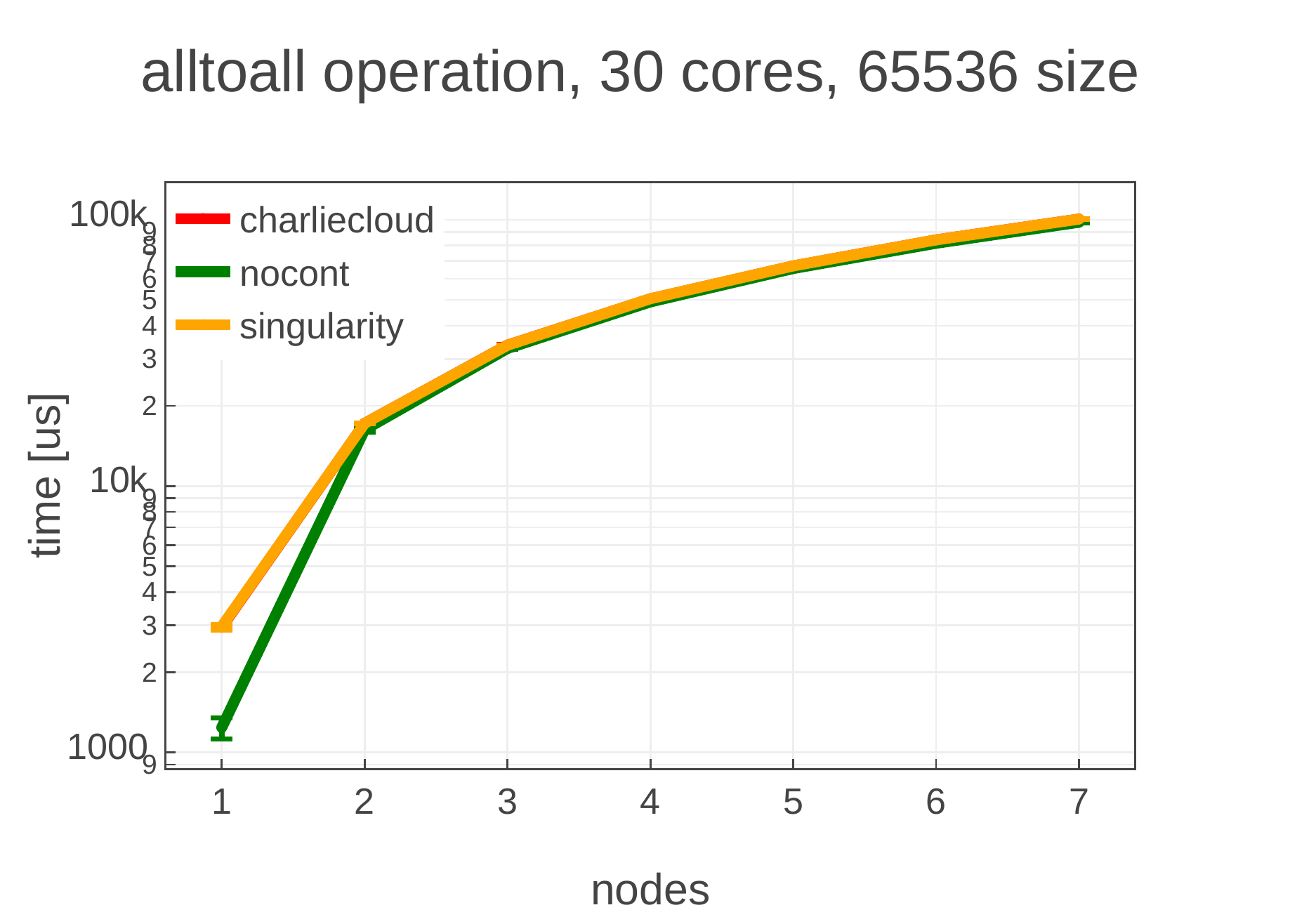}
        \caption{Test \textit{all-to-all} - 65 kB}
        \label{fig:osu_alltoall_65kb_30c}
    \end{subfigure}
    \begin{subfigure}{0.49\textwidth}
        \includegraphics[width=\textwidth, trim={0 0 0 2cm},clip]{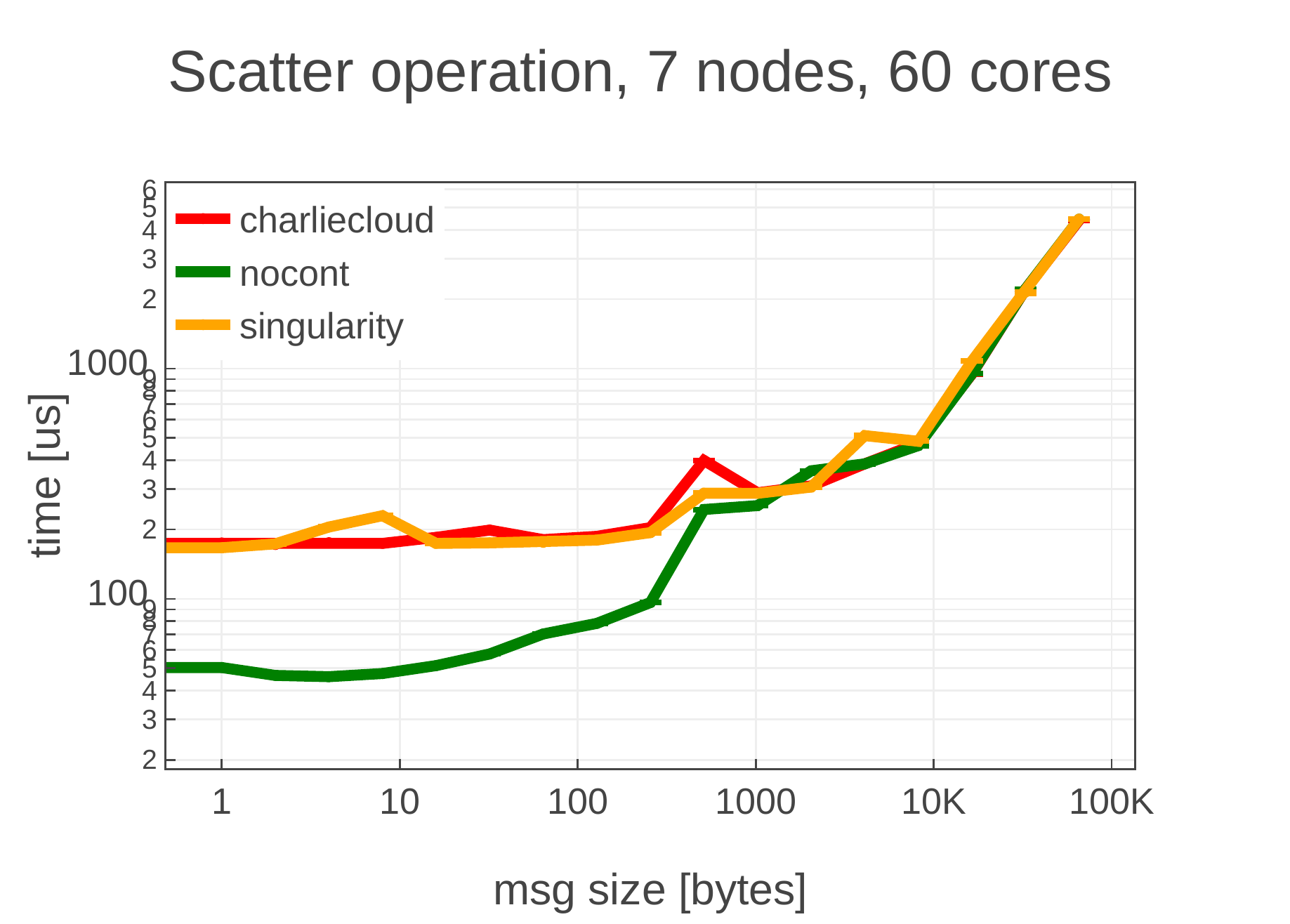}
        \caption{Test \textit{scatter} - 7 nodes}
\label{fig:mpibench_scatter_7nodes_60c}
    \end{subfigure}
    \begin{subfigure}{0.49\textwidth}
        \includegraphics[width=\textwidth, trim={0 0 0 2cm},clip]{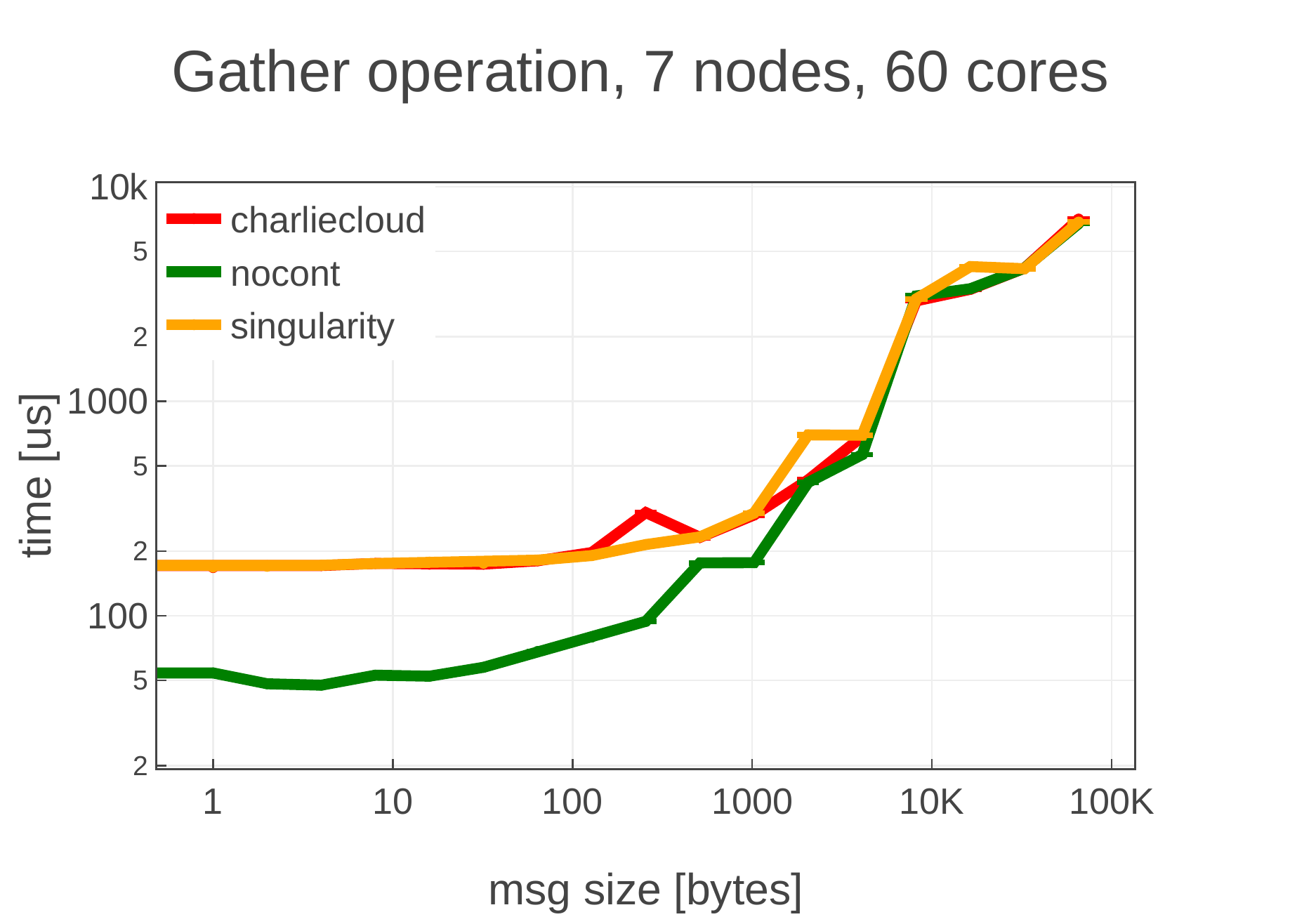}
        \caption{Test \textit{gather} - 7 nodes}
        \label{fig:mpibench_gather_7nodes_60c}
    \end{subfigure}
    \begin{subfigure}{0.49\textwidth}
        \includegraphics[width=\textwidth, trim={0 0 0 2cm},clip]{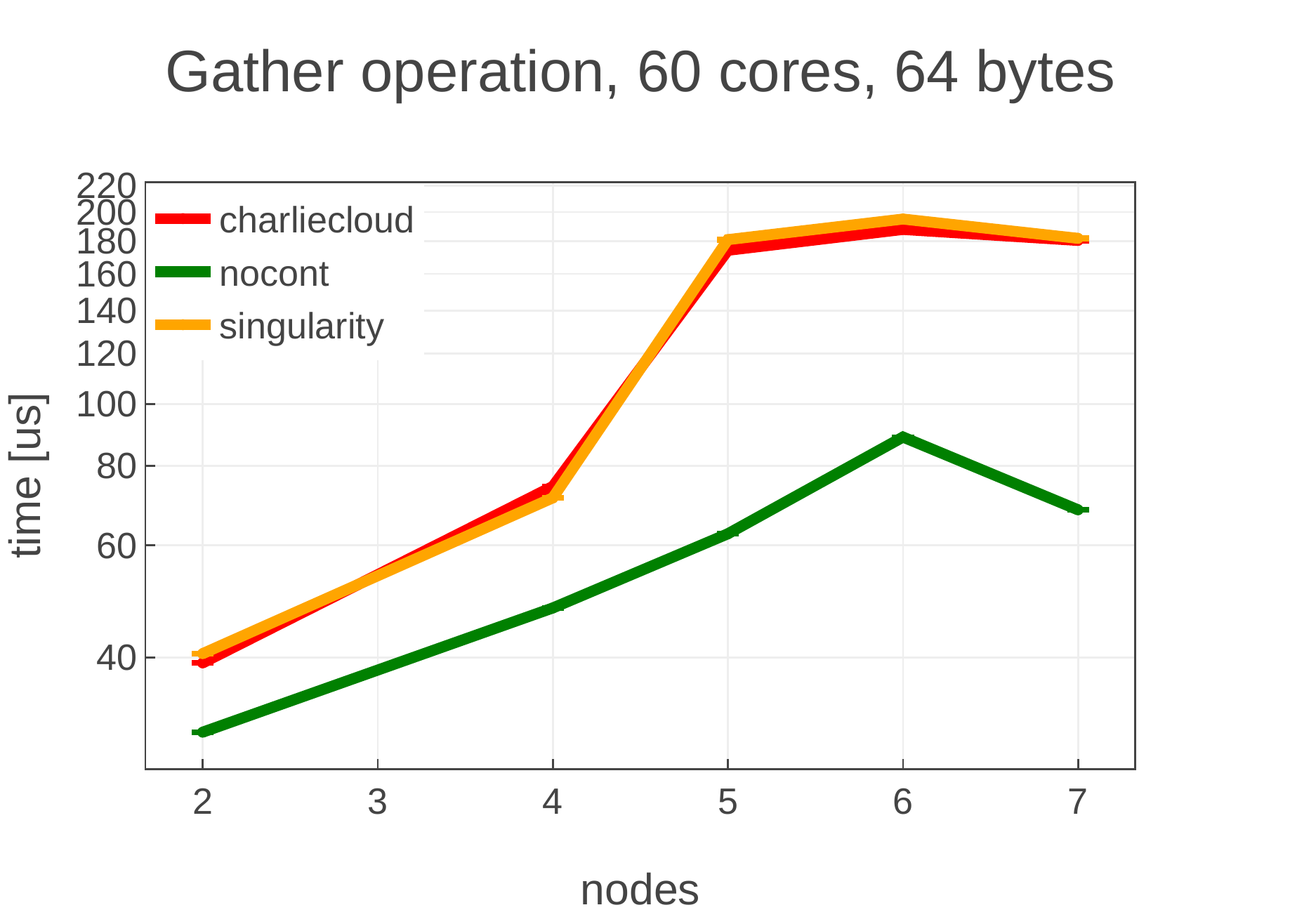} 
        \caption{Test \textit{gather} - 64 bytes}
        \label{fig:mpibench_gather_64bytes_60c}
    \end{subfigure}
    \begin{subfigure}{0.49\textwidth}
        \includegraphics[width=\textwidth, trim={0 0 0 2cm},clip]{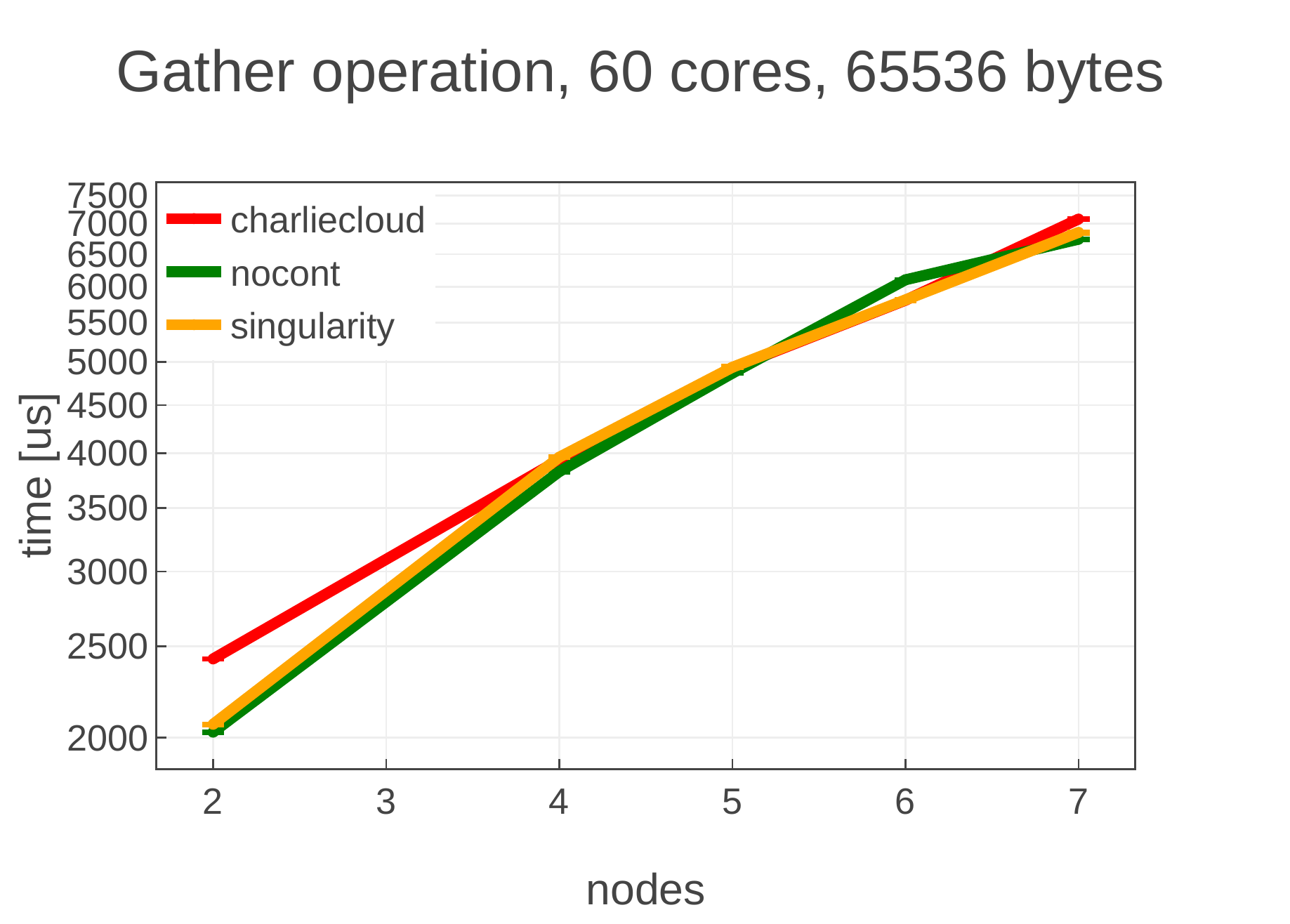}
        \caption{Test \textit{gather} - 65 kB}
    \label{fig:mpibench_gather_65kb_60c}
    \end{subfigure}
    \caption{\blue{MPI experiments}.}
    \label{fig:MPI-exp}
\end{figure}

\blue{In the following, we describe the results of the experiments we carried out. Figure~\ref{fig:MPI-exp} shows some relevant charts that picture different tests run by varying either the cluster or the massage size.}

\blue{Test \textit{latency} was designed to investigate the effect of message size on operation time during the execution of MPI applications with no containers (nocont), Singularity, and Charliecloud. This test was run in two configurations: 2 nodes with 1 core each, and a single node with 2 cores. The message sizes ranged from 1 byte to 4MB, and the operation time was measured in $\mu s$. }

\blue{The results for the two configurations were similar, and the results for the first configuration are shown in Figure~\ref{fig:osu_latency_2nodes_1c}. The chart clearly shows that the three lines (nocont, Singularity, and Charliecloud) are almost overlapped, indicating that the overhead introduced by the container engines in this test was negligible. This suggests that the container engines had a minimal impact on the performance of the MPI applications in ping-pong communications, even if the message size increased. }

\blue{Test \textit{all-to-all} was run with 30 cores per node (leaving half of the computational capacity to process incoming messages), three cluster configurations (2, 4 and 7 nodes), and a variable message size (from 0 to 4MB). All the charts show the average latency in $\mu s$.
Charliecloud and Singularity perform very similarly in all the experiments. When using 7 nodes, the latency is up to 2 times greater than with containers with a message size between 0 and 512 bytes. For larger sizes, the overhead introduced is almost negligible. Similar results are obtained with 2 and 4 nodes with a non-negligible overhead when operating with small message sizes. Although the difference is significant in relative terms, the absolute operation times when dealing with small messages are very small (less than $1 ms$).}

\blue{To better visualize the results, we also run test \textit{all-to-all} with  a fixed message size (either 8, 16 bytes or 65kB) and a variable number of nodes. The overhead does not appear to be strongly correlated with the number of nodes in the system. In the experiment with 8 byte messages the latency introduced by container engines ranges from 1.2 to 2 times the one with no containers (\textit{nocont}); similar values were obtained with 64 bytes (Fig. \ref{fig:osu_alltoall_64_30c}). With 65kb messages, Figure ~\ref{fig:osu_alltoall_65kb_30c}, the overhead is negligible with more than two nodes.}

\blue{These experiments suggest that container engines introduce non-negligible overheads (but yet very small in absolute terms)  with small messages, while it becomes negligible when the size increases (and if more than one node is employed).}

\blue{Tests \textit{scatter/gather} were run with 60 cores per node, three cluster configurations (2, 4 and 7 nodes), and a variable message size (from 0KB to 100KB). All the charts show the average latency in $\mu s$. 
Figures \ref{fig:mpibench_scatter_7nodes_60c} and \ref{fig:mpibench_gather_7nodes_60c} show two runs with 7 nodes for tests \textit{scatter} and \textit{gather}, respectively. Similar results were obtained with 2 and 4 nodes.}

\blue{The obtained results confirm the ones of test \textit{all-to-all}. Charliecloud and Singularity appear to have similar performance and they only introduce overhead with small messages. For example, the two engines added a latency that ranges between 1.3 to 5 times the execution with no container when the message size is between 0 and 256 bytes and we execute \textit{scatter} operations; the overhead becomes negligible with bigger messages. The results obtained with \textit{gather} operations are similar but with an overall smaller maximum overhead.}

\blue{This behavior is even clearer when fixing message sizes (either 8, 64 bytes or 65kB) and varying the number of nodes. Figures     \ref{fig:mpibench_gather_64bytes_60c} and     \ref{fig:mpibench_gather_65kb_60c} show test \textit{gather} with a message size of 64 bytes and 65kB, respectively.
As with test \textit{all-to-all}, the overhead is almost constant and relatively significant with the first two configurations, while with messages of 65kB, the overhead becomes negligible.}

\vspace{0.5cm}
\begin{mdframed}
\textbf{Finding \#\findings.} Container engines dedicated to HPC obtained almost identical performance compared to executions without containers in tests that involved simple MPI communications. When tested with more complex operations, the overhead is still negligible unless the messages exchanged are small. \blue{In this case, the overhead is significant relatively to the executions times which are, in turn, very small (less than $1 ms$) in absolute terms.}
\end{mdframed}

%% file: 6-5._discussion.tex
\blue{The purpose of the second part of our study (Section~\ref{sec-evaluation}) was to investigate the performance of the six container engines. Through our empirical evaluation, we were able to compare the results obtained by these engines across a wide-range of tests and examine how context-specific solutions may or may not outperform general purpose ones in certain metrics}

\blue{In most of the tests, we discovered that HPC-specific solutions outperformed all the other ones. For example,  Charliecloud, Singularity with SIF images, and Sarus obtained the fastest startup times. On the contrary, Podman, balenaEngine and Docker showed noticeable delays. Singularity was by far the fastest in stopping multiple running containers. Shutdown times were generally negligible and constant, except for Podman, which showed a significant variance.}

\blue{Another important aspect we analyzed was the memory usage of the container engine. In HPC contexts, where multiple concurrent processes may be running, it is important to minimize the memory footprint of the container engine to avoid resource contention. In IoT contexts, where resources are often constrained, it is also crucial to minimize the memory usage of the container engine to allow applications to fully utilize available resources. Our experiments showed that Charliecloud resulted in the smallest memory footprint among the six container engines, with no memory being allocated to the container when  idle. Singulariry and Sarus also had a small memory footprint, while Docker obtained the highest memory usage. Although memory is a scarce resource in IoT environments, balenaEngine appears to not optimize its management and obtained performance similar to Docker.}

\blue{The size of container images can be a an important factor to consider, particularly in IoT  where storage may be limited. Our experiments revealed that Charliecloud and Sarus obtained the smallest image sizes among the six container engines. Docker, Podman, and balenaEngine did not offer significant optimization for saving disk space. Singularity uses SIF to compress images, resulting in a significant reduction in disk occupation in some cases.}

\blue{One key factor we considered was the overhead introduced by each container engine. In HPC, where high performance is required, it is essential for the engine to have a minimal overhead. In IoT, where resources are often constrained, it is also important for the container engine to have low overhead to allow applications fully utilizing the available resources. }

\blue{Our experiments revealed, once again, that HPC solutions outperformed all the other engines. In particular, Sarus and Singularity obtained the lowest overhead among the six container engines with performance that are extremely closed to bare-metal ones (maximum 5\% of overhead). Charliecloud also outperformed non-HPC engines. balenaEngine obtained similar performances to Podman, while Docker resulted in the highest overhead (up to 41\%).}

\blue{We also found that the overhead introduced by container engines in HPC-specific tests based on MPI is in general very small and more pronounced for small message sizes. As the message size increased, the overhead decreased and eventually disappeared. This suggests that the overhead introduced by container engines is more significant for fine-grained communication patterns.}

\blue{Overall, we found that balenaEngine had a higher overhead and memory usage as well as larger image sizes on disk compared to other container engines. This may not be suitable for IoT contexts where resources are limited and optimizing for resource usage is important. However, it is important to note that balenaEngine offers a range of features specifically tailored for IoT environments, such as differential image updates to reduce network transfers and on-the-fly image extraction to save disk space. It also provides a range of security features, including an API key for each device and VPN support, which may be beneficial in ensuring the security of IoT systems, as described in depth in Section~\ref{sec-features}.
Similarly, cloud-based solutions may lack some the optimization of HPC-dedicated engines but they provide a broader feature set (especially in terms of orchestration) and they are easier to use and configure.}

\blue{In conclusion, our experiments suggest that there is no one-size-fits-all solution for containers. The best choice of container engine will depend on the specific requirements and constraints of the target environment. However, our results provide some guidance on the trade-offs and potential benefits of the different container engines in each context.}

%% file: 7_related.tex
Containerization emerged as a fundamental enabler for deploying applications in a lightweight and portable way in multiple contexts. For this reason, some significant research work around this technology have been presented in the literature.

Pahl et al.~\cite{Pahl2019} propose a survey on container-based approaches demonstrating the relevance of the technology and its significant impact on the research landscape. 

 Plauth et al.~\cite{plauth_2017_performancesurvey} present a performance comparison of containers, unikernels, whole-system virtualization, native hardware, and combinations thereof. In particular, they assess application performance with HTTP servers and databases and evaluate the startup time, image size, network latency, and memory footprint. Their work compares two container technologies (Docker, LXD), three unikernels (Rumprun, OSv, MirageOS), and two virtualization technology (KVM, Xen). While we share a similar goal with this research, our work is different because we focused only on container engines, we analyze their features (and not only performance), and we compare their performance with different benchmarks in different contexts.

Saha et al.~\cite{Saha_2018_docker} carry out a performance evaluation for executing scientific applications in cloud-based environments with Docker and Singularity. They aim to help practitioners choose the most suitable container engine approach for HPC workloads. They perform four HPC benchmarks (among them also OSU) and find that the performance of different containerization approaches is extremely close to bare metal, a result similar to the one we obtained in our experiments. Another work on this topic is proposed by Arango et al.~\cite{arango2017performance} and analyzes the features of LXC, Docker, and Singularity. According to their results, Singularity containers are the most suitable containers for HPC. Compared to this paper, our work is broader. It describes more and different container engines, their features, and it does not only focus on HPC. 

Liu et al.~\cite{liu2021performance} analyze the performance of containerization in HPC deployments, highlighting how this technology is becoming increasingly important for highly parallel computations. They compare Singularity and Docker against bare-metal executions and they show that for applications that involve intense inter-process communications, containerization provides worse performance compared to bare-metal. In our work, we reach a similar conclusion but only if the communication involves small and frequent messages, while the overhead introduced by containers with larger messages appears to be negligible. As described above, our work does not focus only on HPC and provides a comprehensive comparison among six different container engines.

Potdar et al.~\cite{potdar2020performance} compare the performance of Docker containers and VMs. Their main outcome is that Docker outperforms VMs in every test. Our work is different because compares different container engines with a comprehensive qualitative and quantitative analysis. Similarly, Kozhirbayev et al.~\cite{kozhirbayev2017performance} propose a comparison among Docker, LXC and bare-metal executions. They focus on performance and they discover that, for CPU-bounded tasks, the overhead introduced by containerization is negligible. On the contrary, I/O- and network-intensive applications are faster on bare-metal. Compared to our work, they only compare two container technologies and they focus on their performance, whereas we compared both the features and the performance of six container engines.

Salah et al.~\cite{Salah2017} compare the performance of applications built with the microservice architecture deployed on VM and container on the Amazon Web Service cloud. For container executions, they rely on Elastic Container Service (ECS), a Container-as-a-Service platform provided by Amazon. They discover that the performance of the application running on VMs is significantly better than the ones executed on ECS. Compared to our work, we focus on six container engines and their performance against bare-metal. We also performed some application-level tests (see Section~\ref{subsec-perfovh}) but we did not obtain similar differences in the performance. One possible explanation for this is that ECS introduces additional overhead during the executions.

%% file: 8_conclusion.tex
Containerization is a key enabling technology that increases portability across execution environments, eases application management, and speeds up the scaling and reconfiguration of systems. Container engines are the means to create container images, share them on public or private registries, and execute and manage container instances. 

This paper analyses six container engines, compares their features, and presents some experiments to confront their performance. 
We provide \total{findings} key insights to spread the light on the characteristics of these technologies and on the differences among them. We discovered that, in general, container engines dedicated to HPC are heavily optimized as for performance, cloud-based solutions are richer feature-wise, while IoT tools pose unique challenges that must be tackled by a dedicated approach (e.g., orchestrating a fleet of privacy-sensitive devices). 

This paper focuses on container engines as core enablers for running containerized applications, in the future we will provide an analysis of container orchestrators that are becoming increasingly important for managing single or multiple distributed systems in the Cloud and beyond. 